\documentclass[twocolumn,prb,reprint]{revtex4-2}
\usepackage{amsmath,amssymb} 
\usepackage{amsfonts} 
\usepackage{bm} 
\usepackage{here}
\usepackage{url}
\usepackage{ulem}
\usepackage[colorlinks=true,citecolor=blue,linkcolor=blue,filecolor=blue,urlcolor=blue]{hyperref} 
\usepackage[pdftex]{graphicx} 
\usepackage{braket} 
\usepackage{physics}

\begin{document}
\title{
    Complete Entanglement Structure of the Kitaev Honeycomb Spin Liquid from Exact Tensor Networks
    }
\author{Hidehiro Saito}
\email{saito-hidehiro722@g.ecc.u-tokyo.ac.jp}
\author{Chisa Hotta}
\email{chisa@phys.c.u-tokyo.ac.jp}
\affiliation{Department of Basic Science, University of Tokyo, Meguro-ku, Tokyo 153-8902, Japan}
\date{\today}
\begin{abstract} 
While most of the features of the Kitaev honeycomb model are already clarified from 
the combined ${\mathbb Z}_2$ plaquette and Majorana fermion picture, 
we show that retrieving the exact quantum many-body form gives a clue to fix the remaining issues 
on its spin liquid property. 
We provide an exact two-dimensional tensor network description of a full set of solutions 
from a random product state, 
by sequential quantum-number projection and Gutzwiller projection, 
both expressed as local matrix-product operators of dimension $D=2$. 
The ground and excited states share the same bond dimension yet exhibit distinct Schmidt spectra, 
distinguishing their area-law and volume-law entanglement. 
We demonstrate that the topological entanglement entropy emerges {\it for all eigenstates only after}
the Gutzwiller projection onto the physical spin Hilbert space, 
where the gauge constraint removes exactly half of the Schmidt states. 
This establishes that excited states are likewise topologically ordered. 
We further show that there exists a projection-induced collapse of the wave function into 
a single gauge sector in the Parton mean-field states that is known to give a good description of the spin liquids. 
The demonstrations on the $S=1/2$ kagome and triangular Heisenberg models suggest that the present perspective 
is applied to a wider class of models hosting ${\mathbb Z}_2$ spin liquids.
Our finding visualizes a bulk-boundary correspondence with the Wilson-loop characterization of topological order. 
\end{abstract}
\maketitle
\par
\noindent
\section{Introduction} 
The Kitaev honeycomb model \cite{Kitaev2006} has long served as a canonical platform for realizing a ${\mathbb Z}_2$ quantum spin liquid (QSL), owing to its analytical tractability using classical ${\mathbb Z}_2$ plaquettes and free Majorana fermions. 
In candidate materials like Li$_2$IrO$_3$, Na$_2$IrO$_3$\cite{Chaloupka2010,Singh2010,Singh2012} and $\alpha$-RuCl$_2$\cite{Banerjee2016,Nasu2016}, additional Heisenberg or $\Gamma$ interactions are present 
that make the state no longer exactly solvable\cite{Jackeli2009, Rau2014,Rau2016}. 
Still, the insights from the Kitaev limit help their understanding, 
with Majorana fermions regarded as a hallmark of the QSL\cite{Qiang2024}. 
\par
However, the true QSL wavefunction is not the bare Majorana state but its Gutzwiller projection onto 
the physical Hilbert space. Since this distinction does not explicitly affect most of the non-topological 
thermodynamic properties\cite{Nasu2015}, 
it was not clear whether the post-projected states that are no longer treated analytically, can have more importance. 
Mathematically, for fermionic models with four Majorana modes, 
one can construct a solvable toric-code type Hamiltonian\cite{Kitaev2003} 
that commutes with the loop-forming Gutzwiller projector, 
and this projection is indispensable to have the topological order of that Hamiltonian\cite{Fontana2025}. 
For generic models, however, the projection cannot be carried out analytically, 
making an explicit quantum many-body formulation of the QSL essential to address this issue. 
\par
Unfortunately, only a few works have constructed exact quantum many-body states of the Kitaev honeycomb model. 
Tensor-network (TN) representation of the ground state was successfully obtained 
by encoding the Jordan-Wigner, Fourier, and Bogoliubov transformations into a 3D unitary TN circuit 
of polynomial depth\cite{Orus2017}. 
However, the fermionic operation in such standard TN generally requires a relatively complex algorithm.
\footnote{In Ref.[\onlinecite{Orus2017}], decomposing the operations including 
Jordan-Wigner, Fourier, and Bogoliubov transformations to 2-qubit gates require 
explicit fermionic SWAP gates, which are extra to the standard TN for the spin Hamiltonian. 
Our method also treats fermions, but the fermionic swaps are implicitly encoded in the single-site MPO, 
and requires no extra treatment compared to the projection operation developed in Ref.[\onlinecite{Saito2025}] 
to obtain an exact solution for the frustration-free spin Hamiltonian. 
The computational cost of Ref.[\onlinecite{Orus2017}] is measured by the polynomial depth unitary circuit, 
which is comparable to our $O(N)$ cost. 
The memory cost should also be comparable if one performs the same size calculation without approximation
}
Approximate two-dimensional TN states based on projected entangled pair states (PEPS) have also been 
proposed \cite{Corboz2014}; they break the QSL symmetry while restoring it variationally to some extent, 
and are largely restricted to the ground state\cite{Lee2019}.
The approximation sacrifices the accurate description of the underlying gauge structure, 
e.g., mixes different topological sectors, and raises difficulty in the evaluation of the entanglement spectrum. 
Some works on infinite cylinders have succeeded in separating the topological sectors 
and the entanglement spectrum for the approximate DMRG states\cite{Cincio2013,He2014}.
However, obtaining the numerically exact entanglement spectrum for a sufficiently large system 
remains an utmost challenge. 
\par
In this paper, we propose an exact and simple two-dimensional TN representation 
of the full spectrum of the Kitaev honeycomb model. 
The construction relies on matrix-product-operator (MPO) representations\cite{Verstraete2004,Schollwock2011,Frowis2010} 
of {\it the global projectors associated with plaquette and Majorana fermion quantum numbers}, 
reducible to products of local MPOs with bond dimension $D=2$ and no additional fermionic SWAP gates. 
Starting from a randomized one-dimensional matrix product state (MPS)\cite{Fannes1992, Schollwock2011}, 
we successively apply these projectors while truncating idle bonds, 
and finally perform a Gutzwiller projection from Majorana to spin degrees of freedom.  
Our TN enables efficient local operations, including Majorana braiding, bipartition, and spatially local projections, 
via product-MPO manipulations. 
We find that the topological entanglement entropy(EE) \cite{Kitaev2006-add, Yao2010}, $-\ln 2$, 
appears only after the Gutzwiller projection. 
The projection reduces all the Schmidt states exactly by half, as they allow only one of 
the two mutually exclusive $\mathbb Z_2$-gauge states to survive. 
Similar collapse of the Schmidt states by the Gutzwiller projection  
also occurs in the spin-liquid state of the $S=1/2$ kagome antiferromagnet, 
despite the absence of explicit gauge variables or loop degrees of freedom in its Hamiltonian. 
The underlying global gauge constraint on the Hilbert space is therefore 
an intrinsic property of emergent topological order, 
providing a visible bulk counterpart to the boundary gauge sector characterized by Wilson loops
\cite{Levin2005,Levin2006}. 
\par
\begin{figure}[tbp]
   \centering
   \includegraphics[width=8.5cm]{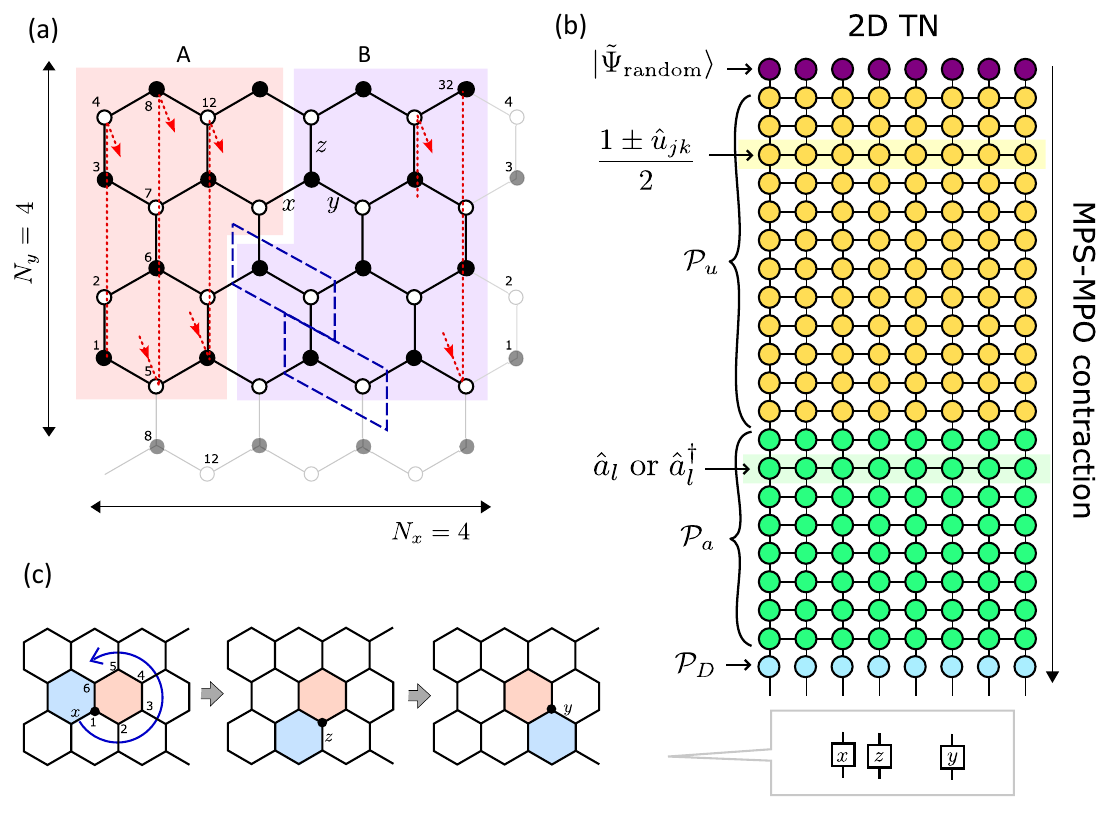}
    \caption{
        (a) Finite-size cluster of a Kitaev honeycomb model ($N=32$ with $N_x=N_y=4$) 
         with bond labels, $\gamma=x,y,z$, and with PBC. 
         We construct the MPS along the 1D path shown in a broken line. 
        (b) Graphical representation of the TN that provides the exact eigenstate of the Kitaev honeycomb model. 
         The first row shows the MPS of $|\tilde \Psi_{\rm random}\rangle$. 
         Operating $(1\pm \hat{u}_{jk})/2$ on successive rows, totally $3N/2$, 
         and then operating $N$ rows of MPO of either $\hat{a}_l$ or $\hat{a}_l^\dag$, 
         we obtain the fermionic $|\tilde \Psi^{M}_{\bm u,\bm n}\rangle$. 
         Finally $N$-single operators, ${\cal P}_D$, project it onto $|\Psi_{\bm u,\bm n}\rangle$.
        The neighboring MPOs and MPS are contracted for each operation. 
        (c) Braiding process, where the plaquette $\alpha$ goes around $\beta$ by 
        $\sigma^y_6\sigma^z_5\sigma^x_4\sigma^y_3\sigma^z_2\sigma^x_1$, operating independently on six 
        tensors around the $\beta$ plaquette. 
    }
\label{f1}
\end{figure}
\noindent
\section{Formulation} 
\subsection{Preliminaries} 
We consider a spin-1/2 Kitaev honeycomb model on a $N=2N_x\times N_y$ lattice (Fig.~\ref{f1}(a)) 
with a periodic boundary condition(PBC) given as 
\begin{equation}
{\cal H}= - \sum_{\langle j k \rangle^\gamma} J_{\gamma}\hat \sigma^\gamma_j \hat\sigma^\gamma_k,
     \label{eq:ham}
\end{equation}
where $\langle jk \rangle^\gamma$ runs over all nearest neighbor sites, and 
$\gamma=x,y,z$ denotes either of the three different directions the $jk$-bond is pointing. 
The up/down spin degrees of freedom are extended to the representation of four Majorana fermions, 
$\hat b_i^x,\hat b_i^y,\hat b_i^z,\hat c_i$, where $\hat\sigma_i^\alpha = i \hat{b}^\alpha_i \hat c_i$. 
For the $\langle jk \rangle^\gamma$ neighbors, these Majoranas form 
a bond operator given as, $\hat{u}_{jk}=i\hat b_j^{\gamma} \hat b_k^{\gamma}$, 
and by using them, Eq.(\ref{eq:ham}) is rewritten in the quadratic form, 
\begin{equation}
  {\cal H}_{M} = i \sum_{\langle j k \rangle^\gamma}  J_{\gamma} \hat u_{jk} \hat c_j \hat c_k. 
    \label{eq:fer_ham} 
\end{equation}
Here, because $[{\cal H}_{M},\hat{u}_{jk}]=0$, 
and $\hat{u}_{jk}$ has the eigenvalue $u_{jk}=\pm 1$, 
this Majorana Hamiltonian is block diagonalized into the subspace specified by 
a set of quantum numbers $\bm u=\{u_{jk}\}$ over $2^{3N/2}$ different combinations. 
Suppose we have one such set, and the resultant matrix representation 
of the Hamiltonian on that subspace is given in the $N\times N$ form, 
$iA = \{2iJ_{\gamma} u_{jk}\}_{j,k=1}^N$, with its elements satisfying $u_{jk}=-u_{kj}$. 
Diagonalizing it as $iA=QEQ^\dagger$, we find, 
${\cal H}_{M} =\sum_{\ell=1}^{N/2} \epsilon_\ell (\hat{a}_\ell^\dag \hat{a}_\ell-1/2) $, 
where $\hat{a}_\ell=1/\sqrt{2}\sum_{j=1}^N Q_{j\ell} \hat c_j$ 
satisfies $\hat{a}_\ell=\hat{a}_{N+1-\ell}^\dagger$ when ${\rm diag}E=\{-\epsilon_\ell\}_{l=1}^N$ is arranged in ascending order.\par
The Majorana fermion number operator is, $\hat n_\ell= \hat{a}_\ell^\dag \hat{a}_\ell$, where $n_\ell=0$ or 1. 
Therefore, each eigenstate $|\tilde{\Psi}_{\bm u,\bm n}^{M} \rangle$ of ${\cal H}_{M}$ 
has the quantum numbers, $\bm u=\{u_{jk}\}$ and $\bm n=\{n_\ell\}$. 
To put these eigenstates back to those of the physical Hamiltonian, 
the operator, $\hat{D}_i=\hat b_i^x \hat b_i^y \hat b_i^z \hat c_i$, whose eigenvalue is $\pm 1$, 
is useful, as the sector $D_i=1$ forms a subspace of spin degrees of freedom on site $i$. 
Using their products, the fermionic wave function 
is projected onto the original Hilbert space as 
\begin{equation}
   |\Psi_{\bm u,\bm n} \rangle \propto {\cal P}_D|\tilde{\Psi}_{\bm u,\bm n}^{M} \rangle, 
\quad  {\cal P}_D = \prod_{i=1}^N \frac{1+\hat{D}_i}{2}. 
    \label{eq:pd}
\end{equation}
These preliminary formulas are well established \cite{Kitaev2006}. 
\vspace{3mm}
\\
\subsection{Exact tensor network} 
For a given values of quantum numbers, $\bm u=\{u_{jk}\}=\pm 1$ and $\bm n=\{ n_\ell\}=$ 0 or 1, 
there are corresponding projectors given in the form, 
\begin{align}
{\cal P}_u &= \prod_{\langle jk \rangle^\gamma } \hat p_{jk},\quad 
\hat p_{jk}=\frac{1\pm \hat u_{jk}}{2},\; (j<k, \;u_{jk}=\pm 1)
    \label{eq:pu} \\
{\cal P}_a &= \prod_{\ell =1}^{N/2}  \hat q_\ell, \quad \quad
\hat q_\ell=\left\{\begin{array}{ll}
    \hat{a}_\ell^\dag \hat{a}_\ell & (n_\ell=1) \\
    \hat{a}_\ell \hat{a}_\ell^\dagger & (n_\ell=0) 
    \end{array}\right. , 
    \label{eq:pn} 
\end{align}
where for $j>k$, $\hat p_{jk}=(1\mp \hat u_{kj})/2$. 
Previously, fermionic MPS-MPO representations with a graphical structure similar to Fig.~\ref{f1}(b) 
were developed as variational ansatz for ground states~\cite{Jin2020,Jin2021,Jin2025}. 
Our framework differs conceptually: rather than constructing the target state by applying fermionic 
creation operators to the vacuum\cite{Wu2020}, 
we start from a maximally informative random state and extract 
the target Hilbert subspace by projection. This philosophical difference 
is analogous to the contrast between METTS~\cite{White2009} 
and TPQ~\cite{Iwaki2021} for thermal states, where the former builds a mixed state 
as an ensemble of initially unentangled product states by minimally developing entanglement, 
whereas the latter projects a highly entangled random state into a single pure state.
\par
Exact eigenstates are obtained because the projectors $\hat p_{jk}$ and $\hat q_\ell$ admit 
{\it a factorized MPO representation, with each local tensor acting only on a single-site Hilbert space. } 
We construct such a representation using MPOs acting on an MPS wavefunction defined on a finite cluster with PBC. 
The MPS is defined along a helical path covering the two-dimensional lattice, 
as illustrated in Fig.~\ref{f1}(a). 
For the single-site fermionic basis of dimension $d=4$ defined as 
$|0\rangle, |\uparrow\rangle, |\downarrow\rangle, 
|\uparrow \downarrow\rangle$ \cite{Keller2015, McCulloch2017}, 
the MPO of $\hat p_{jk}$ and $\hat q_\ell$ acting on the $\gamma$-th bond and $\ell$-th site 
can be represented with bond dimension $D=2$ as 
\begin{align}
&\frac{1 \pm \hat{u}_{jk}}{2}  
    =\; I_1\otimes \cdots \otimes I_{j-1} \otimes
    \begin{bmatrix}
        \frac{1}{2}I_j & \pm \frac{i}{2} b_j^{\gamma} \tilde{F}_j^{\rm (i)}
    \end{bmatrix} \otimes 
    \notag \\
    &\hspace{10mm}\begin{bmatrix}
        I_{j+1} & 0 \\ 0 & F_{j+1}
    \end{bmatrix} \otimes \cdots \otimes
    \begin{bmatrix}
        I_{k-1} & 0 \\ 0 & F_{k-1}
    \end{bmatrix} \otimes 
    \begin{bmatrix}
        I_k \\ \tilde{F}_k^{\rm (f)} b_k^\gamma
    \end{bmatrix}
    \notag \\
    &\hspace{10mm} \otimes I_{k+1} \otimes \cdots \otimes I_N,
    \label{eq:pu_mpo} \\
 &\hat{a}_\ell 
    =\; 
    \begin{bmatrix}
        \frac{1}{\sqrt{2}} F_1 & \frac{Q_{1l}}{\sqrt{2}} F_{\uparrow 1} c_1
    \end{bmatrix} \otimes 
    \begin{bmatrix}
        F_2 & Q_{2l} F_{\uparrow 2} c_2 \\ 0 & I_2
    \end{bmatrix} \otimes \cdots \otimes
    \notag \\
    &\hspace{8mm}\begin{bmatrix}
        F_{N-1} & Q_{N-1 \;l} F_{\uparrow N-1} c_{N-1} \\ 0 & I_{N-1}
    \end{bmatrix} \otimes 
    \begin{bmatrix}
        Q_{N l} F_{\uparrow N} c_{N} \\ I_N
    \end{bmatrix},
    \label{eq:a_mpo}
\end{align}
where $F_{\uparrow}= \textrm{diag}(1,-1 ,1,-1)$, $F_{\downarrow} = \textrm{diag}(1,1 ,-1,-1)$, and 
$F_j=F_\uparrow F_\downarrow $ is the fermion string required by the anticommutation relation, 
$\tilde{F}_j^{\rm (i)}= F_j$ ($\gamma=x,y$) and $F_{\downarrow}$ ($\gamma=z$), 
and $\tilde{F}_k^{\rm (f)}=I_k$  ($\gamma=x,y$) and $F_{\uparrow}$ ($\gamma=z$). 
\par
In addition to the local MPO, the key to obtaining a numerically exact MPS eigenstate 
is to start from an initial random product state $|\tilde{\Psi}_{\rm random} \rangle$, 
where $\chi=1$ turns out sufficient. 
By operating Eqs.(\ref{eq:pu}) and (\ref{eq:pn}), we obtain the exact TN eigenstate as 
\begin{equation}
    |\tilde{\Psi}^M_{\bm u,\bm n} \rangle 
     = {\cal P}_a  {\cal P}_u |\tilde{\Psi}_{\rm random} \rangle, 
    \label{eq:eig_fer}
\end{equation}
which is graphically illustrated in Fig.~\ref{f1}(b). 
At each step of applying $\hat p_{jk}$ or $\hat q_\ell$, we Schmidt decompose the state and discard the idle bond dimension. 
The local contraction is allowed because Eq.(\ref{eq:pu_mpo}) is a product of local operators. 
This process does not imply an approximation 
\footnote{The numerical accuracy of the ground state and randomly chosen excited state 
at $N=32$ when starting from the bond dimension $D=1$ are 
$|\langle {\cal  H} \rangle -E_{\rm exact}|= 2.85\times 10^{-10}$ and $2.66\times 10^{-15}$, respectively,  
against the analytical exact solutions $E_{\rm exact}$.}, 
unlike the standard variational usage of MPS that uses finite $\chi$ by discarding small Schmidt states. 
Finally, on the last row of the figure, the Gutzwiller single-site MPO, Eq.(\ref{eq:pd}), is applied 
to make it the eigenstate of Eq.(\ref{eq:ham}). 
\par
Before performing these processes, we need to choose $\bm u$ and $\bm n$ 
to satisfy the conservation laws; 
There are $N/2+2$ loop operators commuting with the Hamiltonian: 
two global loop operators, $W_x, W_y$, winding around torus, 
and $N/2$ plaquette operators $\hat{W}_p=\prod_{j\in {\cal L}_p} \sigma_j^{\alpha_j}$, 
where $\alpha_j=x,y,z$ denotes the bond connected to site $j$ 
but not included in the loop. Considering the constraint, $\prod_{p=1}^{N/2}\hat W_p=\hat 1$, 
there are $N/2+1$ independent conserved $\mathbb{Z}2$ quantum numbers, 
${w_p,w_x,w_y}$, which determine $\bm u$. 
Furthermore, $\bm u$ and $\bm n$ must satisfy the projection constraint 
$\hat D=\prod_{i=1}^{N}\hat D_i=\hat 1$\cite{Yao2009,Yao2010,Willans2011,Loss2011}. 
Details are provided in Appendix \ref{app:tn}. 

\section{Numerical representativity} 
We numerically perform the above calculation on the $N=2 N_x\times N_y=32$ site cluster, 
and test the representativity of our TN. 
\subsection{Braiding}
The prominent advantage of our TN is to easily obtain and manipulate any of the excited states. 
The braiding operation is local and simple (Fig.~\ref{f1}(c)). 
Let the initial state $|\Psi^{(i)}[\alpha,\beta]\rangle$ has $\{w_p,w_x,w_y\}=\{1,1,1\}$, 
except for two colored hexagons, $w_\alpha=w_\beta=-1$. 
We now move the flux on $\alpha$ around $\beta$, which is done by 
operating $\sigma_i^{\gamma}$ with $\gamma$ the bond label that the moving plaquette crosses. 
As it is a single-site operation to our TN on six different sites, the final state, 
$|\Psi^{(f)}[\alpha,\beta]\rangle=
\sigma_6^y\sigma_5^z \sigma_4^x \sigma_3^y\sigma_2^z \sigma_1^x |\Psi^{(i)}[\alpha,\beta]\rangle$, 
is immediately obtained and shows $\langle \Psi^{(i)}[\alpha,\beta]|\Psi^{(f)}[\alpha,\beta]\rangle=W_\beta=-1$. 
\vspace{1mm}
\begin{figure*}[tbp]
   \centering
   \includegraphics[width=17.5cm]{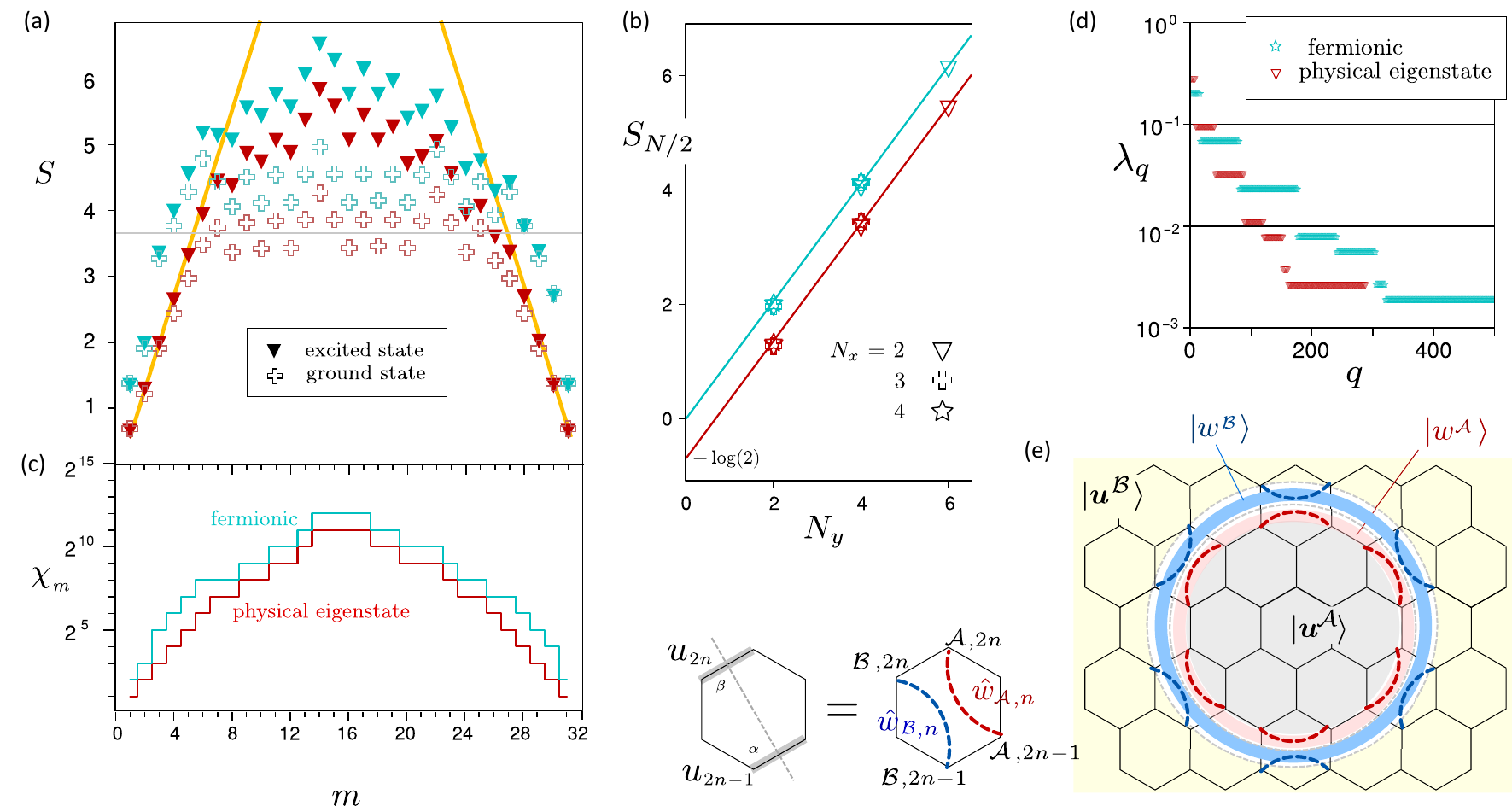}
    \caption{
        (a) Entanglement entropy (EE) $S$ of the exact ground state 
         and the excited state (with $\bm u$, $\bm n$ randomly chosen) at $J_x=J_y=J_z=1$ 
        for the bipartition $m$ and $N-m$ of the $N=32$ cluster in panel (a). 
         The blue and red lines/symbols show the results of the fermionic and physical eigenstates, respectively. 
    (b) Size scaling of $S_{N/2}$ for several system sizes. 
    (c) Bond dimension, $\chi_m$, at the $m$th bond from the left edge. 
    (d) The largest 500 Schmidt values for the bipartition of the ground state 
     fermionic $|\tilde\Psi^M_{\bm u,\bm v}\rangle$ and physical states $|\Psi_{\bm u,\bm v}\rangle$ 
     before and after the Gutzwiller projection, respectively (see also Appendix \ref{sec:dist}).
    (e) Bipartition to $|\psi^{\cal A}\rangle$ and $|\psi^{\cal B}\rangle$, with $|w^{\cal R}\rangle$ forming boundary states 
    belonging to $\cal R=$A, B. 
    }
    \label{f2}
\end{figure*}
%
\begin{figure*}[tbp]
   \centering
   \includegraphics[width=17.5cm]{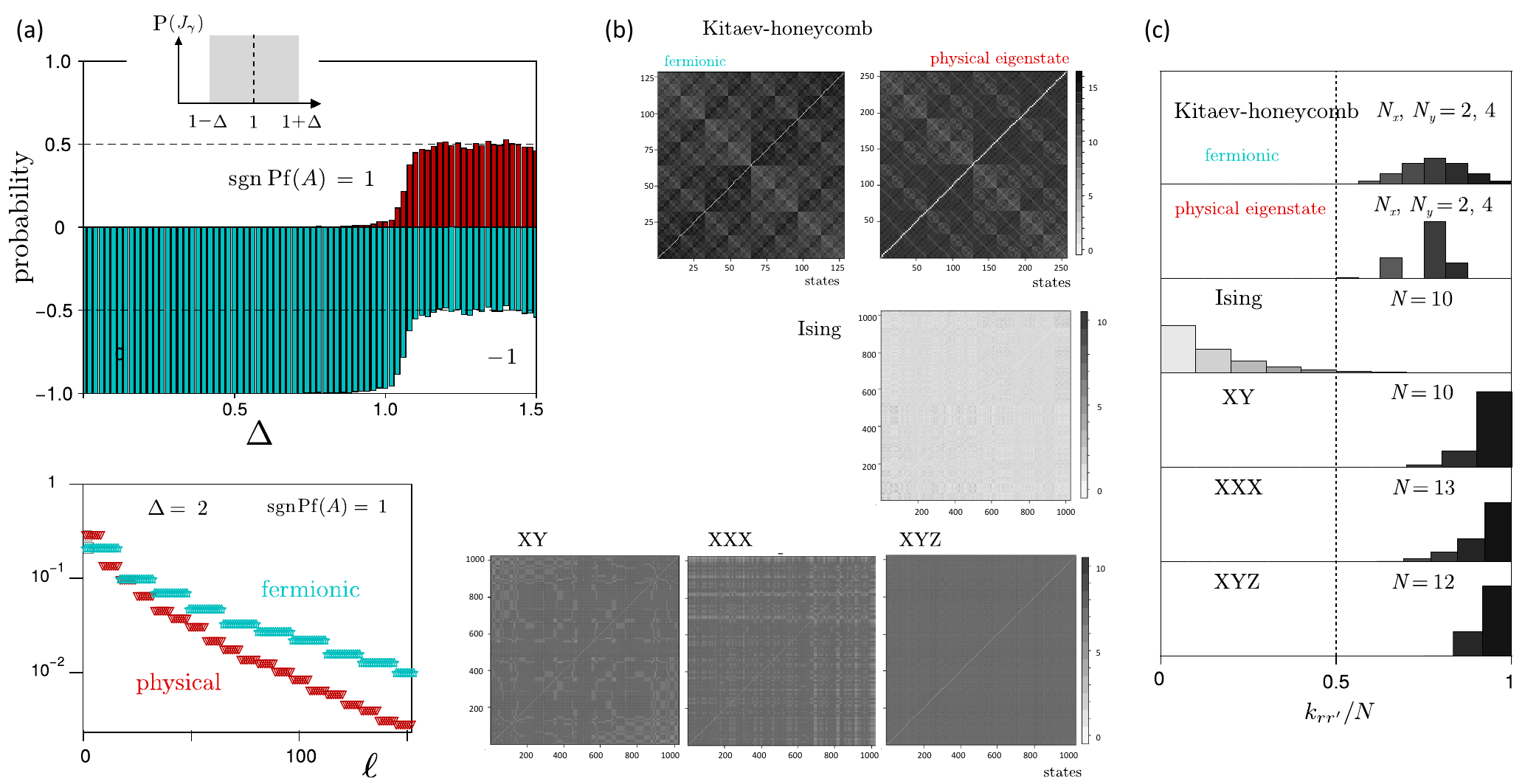}
    \caption{ (a) Signed fraction of Pfaffian in $D$ for the bond-random $J_\gamma$ Kitaev honeycomb model 
    with $N_x=N_y=16$ and uniformly random probability distribution $P(J_\gamma)$, in a flux-free sector,  
    where for each given $\Delta$ we take 1000 samples. 
    Inset: $\lambda_\ell$ for $N_{\cal A}=N/2$ at $\Delta=2$ and ${\rm sgn}({\rm Pf}(A))=1$ state 
    with $u_{ij}=-1$ for $J_{ij}<0$. 
   (b)  Density plot of the comparison of two choices of states labeled by $r$ and $r'$ 
    of the Kitaev honeycomb model for fermionic eigenstates and the physical eigenstates 
    with $N_x=2,N_y=4$ obtained by our TN. 
    We choose the random flux sector ($\bm u$ chosen randomly). 
    A similar comparison is made for the spin chain for $N=10-13$ using exact diagonalization 
    for the Ising, XX, Heisenberg (XXX), and XYZ models for the full eigenstates. 
    (c) Histogram of the value of $k_{rr'}$ (range of the minimum overlap of Schmidt bases between the two eigenstates)
    of different models. 
    }
\label{f3}
\end{figure*}
\subsection{Entanglement entropy} 
Figure \ref{f2}(a) shows the entanglement entropy (EE), $S$, about the bipartition $m$ to $N-m$ 
(see Fig.~\ref{f1}(a)) 
for the ground state and the excited state, where for the latter, 
we take the randomly chosen sector, $\bm u,\bm n$. 
The excited state shows a typical volume law behavior (Appendix \ref{sec:volumelaw}), 
and the ground state exhibits an area-law plateau. 
We test in Fig.~\ref{f2}(b) the size scaling of $S_{N/2}$ for both the fermionic $|\tilde{\Psi}^{M}_{u,n} \rangle$ 
and physical state $|{\Psi}_{u,n} \rangle$. 
Both cases are exactly linear-in-$N_y$ with equal slopes for different choices of $N_x=2,3,4$. 
Notably, the physical ground state has a finite topological term, $-\ln 2$, 
that identifies the ${\mathbb Z}_2$ QSL, whereas the intercept is zero for a fermionic eigenstate. 
The same reduction of EE by $-\ln 2$ after projection appears for the {\it excited states}. 
Interestingly, we find that when ${\cal P}_D$ is applied to $|\tilde{\Psi}^{M}_{u,n} \rangle$, 
the number of Schmidt states at the bipartition is {\it exactly reduced by half 
for each Schmidt value} and the value $\lambda_q$ is scaled by $\sqrt{2}$ instead
(Fig.~\ref{f2}(d)). 
This means that the whole solution space is reduced simultaneously by half, 
and this applies to all eigenstates. 
\par 
As all the solutions are exact, the number of nonzero Schmidt values is equal to the bond dimension $\chi_m$. 
Figure~\ref{f2}(c) shows $\chi_m$ as functions of $m$, where we find that 
they are exactly the same for all eigenstates. 
The maximum value at the system center, 
is given as $\chi_{\rm max}=2^{(N_x+2)N_y/2}, 2^{(N_x+2)N_y/2-1}$, respectively, 
for the fermionic and physical states. 
In general, $\chi_{\rm max}$ grows in power and exponential of system size 
for the states with area law and volume law EE, respectively. 
However, our $\chi_{\rm max}$ has the same size dependence, 
regardless of being a ground state or excited states, 
which is because their construction is formally the same. 
It therefore appears that the area and volume law difference comes from the distribution of Schmidt values 
within the same $\chi_{\rm max}$. 
\par
The big advantage of our TN description is that the Schmidt states are available. 
With this, one can show that the post-projection states with different quantum numbers do not fully 
share the Schmidt state, or in other words, the Majorana fermions are quasi-local. 
(See Appendix \ref{sec:localmajorana} for mathematical proof). 
\subsection{Topological order} 
\label{sec:topo}
\subsubsection{Schmidt values} 
\label{sec:topo-schmidt}
Following our numerical observation on the reduction of the Schmidt states, 
we analytically show that the constraint $D_i=1$  supported by ${\cal P}_D$ creates a topological term of EE. 
Let us first rewrite the fermionic eigenstate using the $\mathbb Z_2$ bond gauge $|\bm u\rangle$ and fermionic part 
$|M_{\bm u}\rangle$ explicitly as $|\tilde \Psi^M_{\bm u,\bm v}\rangle=|\bm u\rangle|M_{\bm u}\rangle$. 
As we bipartite it to ${\cal A}$ and ${\cal B}$ along the closed loop, which cuts totally $2L$ different $\mathbb Z_2$ bonds 
(Fig.~\ref{f2}(e)), 
the $2L$ different bond gauges are rearranged in a way that they 
reconnect and form boundary gauge operators $\hat w_{{\cal A},n}=i \hat b^\alpha_{{\cal A},2n-1} \hat b^\beta_{{\cal A},2n}$ and 
$\hat w_{{\cal B},n}=i \hat b^\alpha_{{\cal B},2n-1} \hat b^\beta_{{\cal B},2n}$, 
whose eigenvalues rely on $|u_{2n-1},u_{2n}\rangle=|\pm,1\pm 1\rangle$ and $|\mp 1,\pm 1\rangle$ on the cut as 
$|w_{{\cal A},n}, w_{{\cal B},n}\rangle=(|-1,1\rangle \pm |1,-1\rangle )/\sqrt{2}$ 
and $(|1,1\rangle \pm |-1,-1\rangle)/\sqrt{2}$, 
each confined inside the region ${\cal A}$ or ${\cal B}$. 
Accordingly, 
the gauge states are divided as $|\bm u\rangle=|\bm u^{\cal A}\rangle |w^{\cal A}\rangle |\bm u^{\cal B}\rangle |w^{\cal B}\rangle$,  where $|\bm u^{\cal R}\rangle$ are the ones inside region ${\cal R}={\cal A},{\cal B}$ that do not touch the cut, 
and $|w^{\cal R}\rangle$ are the boundary gauge states; these facts are known so far\cite{Yao2010}. 
Since $|\tilde\Psi^M_{\bm u,\bm v}\rangle$ is the product of gauge and matter part, 
the latter is trivially Schmidt decomposed 
into $|M^{\cal A}_{\bm u \ell}\rangle|M^{\cal B}_{\bm u \ell}\rangle$ and we obtain an explicit Schmidt form, 
\begin{align}
|\tilde \Psi^M_{\bm u,\bm v}\rangle&= 
\sum_{\ell=1}^\chi \lambda_\ell |M^{\cal A}_{\bm u \ell}\rangle|M^{\cal B}_{\bm u \ell}\rangle
\!\!\!\sum_{\{w_n=\pm 1\}} \!\!|\bm u^{\cal A}\rangle |\bm u^{\cal B}\rangle 
\frac{(-)^\gamma}{\sqrt{2^{L}}} |w^{\cal A}\rangle |w^{\cal B}\rangle
\notag \\
& = \sum_{\ell=1}^\chi \sum_{\{w_n=\pm 1\}} (-)^\gamma \frac{\lambda_\ell}{\sqrt{2^{L}}} 
|\psi^{\cal A}_{\ell}\rangle |\psi^{\cal B}_{\ell}\rangle, \;
\label{eq:schmidt}
\end{align}
with $|\psi^{\cal R}_\ell\rangle= |\bm u^{\cal R}\rangle |w^{\cal R}\rangle |M^{\cal R}_{\bm u \ell}\rangle$, 
and some sign $(-)^\gamma$, 
where each $\lambda_\ell$ has $2^{L}$ degeneracy coming from 
the $w_n=\pm 1$ factor that appear in choosing  $|w_{{\cal A},n},w_{{\cal B},n}\rangle$. 
Let us then divide the global gauge operator as, $\hat D=\hat D_{\cal A} \cdot \hat D_{\cal B}$. 
Importantly, we find that $|\psi^{\cal R}_{\ell}\rangle$ is the eigenstate of $\hat D_{\cal R}$ 
(see Appendix \ref{app:schmidt} Eq.(\ref{eq:DAeig}) for proof). 
\par
The Gutzwiller projector is separated into the product, 
${\cal P}_D={\cal P}_D^{\cal A}{\cal P}_D^{\cal B}$ 
which allows us to project 
the Schmidt states independently for each $\ell$ as ${\cal P}_D^{\cal R}|\psi^{\cal R}_{\ell}\rangle$. 
Accordingly, the Gutzwiller projected eigenstates of the spin Hamiltonian are written explicitly 
by taking account of the normalization factor $\sqrt{2^{N-1}}$ of the wave function as 
\begin{align}
|\Psi_{\bm u,\bm v}\rangle &= \sqrt{2^{N-1}} {\cal P}_D |\tilde \Psi^M_{\bm u,\bm v}\rangle 
= \sqrt{2^{N-1}}{\cal P}_D^{\cal A} {\cal P}_D^{\cal B} | \tilde \Psi^M_{\bm u,\bm v}\rangle 
\notag \\
&= \sqrt{2^{N-1}} \sum_{\ell=1}^\chi
   \sum_{\{w_{\cal A}\},D_{\cal A}=1}
   \frac{\lambda_\ell}{\sqrt{2^{L}}} 
     \big( {\cal P}_D^{\cal A} |\psi^{\cal A}_{\ell}\rangle\big) \big({\cal P}_D^{\cal B} |\psi^{\cal B}_{\ell}\rangle\big) 
\notag \\
& = \sum_{\ell=1}^\chi  \sum_{\{w_{\cal A}\},D_{\cal A}=1}
   \frac{\lambda_\ell}{\sqrt{2^{L-1}}} 
     \Big(\sqrt{2^{N_{\cal A}-1}}\: {\cal P}_D^{\cal A} |\psi^{\cal A}_{\ell}\rangle \Big)
\notag\\
& \quad \quad \quad   \otimes \Big(\sqrt{2^{N_{\cal B}-1}}\: {\cal P}_D^{\cal B}|\psi^{\cal B}_{\ell}\rangle\Big). 
\label{eq:Schmidtall}
\end{align}
Here, $|\psi^{\cal A}_{\ell}\rangle$ is an eigenstate of $D_{\cal A}$ 
as we explain in Eq.(\ref{eq:DAeig}) in Appendix \ref{app:schmidt}. 
Since $|\tilde \Psi^M_{\bm u,\bm v}\rangle$ has $\hat D=1$, there are two choices, $D_{\cal A}=D_{\cal B}=\pm 1$, 
and the Gutzwiller projection enforces $D_i=1$ for all $i$. 
Among $\big( {\cal P}_D^{\cal A} |\psi^{\cal A}_{\ell}\rangle \big)\big({\cal P}_D^{\cal B} |\psi^{\cal B}_{\ell}\rangle\big)$ 
states that include both $D_{\cal A}=D_{\cal B}=\pm 1$ eigenstates, only the $+1$ sector remains, 
which is exactly half of 
$({\cal P}_D^{\cal A}|\psi^{\cal A}_{\ell}\rangle)({\cal P}_D^{\cal B}|\psi^{\cal B}_{\ell}\rangle)$ 
for each $\ell$. 
Consequently, the $2^{L-1}$ degeneracy remains in the projected Schmidt spectrum, 
with each $\lambda_\ell$ multiplied by $\sqrt{2}$, providing an analytical proof of Fig.~\ref{f2}(d). 
Previously, the explicit form of the EE was obtained by Yao and Qi as $S=(L-1)\ln 2 + \alpha L$, 
by calculating the trace of the local density matrix using a Replica trick
\footnote{
In Ref.[\onlinecite{Yao2010}], they directly constructed the density operator $\cal \rho_{\cal A}$ 
of the post-projected state consisting of equal-weight superposition of different gauge configurations. 
The form of $\cal \rho_{\cal A}$ includes the nontrivial off-diagonal 
contributions between different gauge configurations that seem highly nontrivial to 
reconcile with the Schmidt decomposition. 
Instead, they directly evaluated ${\rm Tr}_{\cal A}\big((\rho_{\cal A})^n\big)$, integrated over the regions ${\cal A}$, 
and applied the Replica trick to obtain the von Neumann entropy. 
See Appendix \ref{app:schmidt} for details. 
}.
Here, we disclosed the complete entanglement structure under the gauge constraint, namely, 
at the quantum many-body level for spin Hamiltonian, 
and showed how it generates $-\ln 2$. 
There is another complementary study by D\'ora and Moessner\cite{Dora2018} 
that confines itself to the quantum many-body spin Hamiltonian, 
or equivalently to its Jordan-Wigner fermionic form, successfully separating the gauge and matter sectors and integrating 
the latter, showing that the zero-flux sector of the ground state alone could 
generate a $-\ln 2$ term, which gives a good connection to the cylinder DMRG. 

\subsubsection{Gauge constraint}
With in mind that the $D=1$ constraint to the Majorana fermions plays a crucial role 
in generating $-\ln 2$ at post-projection, 
we reexamine the formula \cite{Yao2009,Willans2011,Loss2011}, 
\begin{equation}
D=(-)^{N_x+N/2} {\rm sgn}\big({\rm Pf}(A)\big) 
\prod_{\langle j,k\rangle} u_{j,k} 
(-)^{\sum_{\ell=1}^{N/2} \hat n_\ell}, 
\vspace{-2mm} 
\label{eq:z2eq}
\end{equation}
by expressing it using the sign of the Pfaffian. 
To understand its implication, we consider a bond-randomized version of Eq.(\ref{eq:ham}) 
with $J_\gamma =  1 + \Delta_{jk}$, where $\Delta_{jk}$ follows the uniformly random distribution at 
$[-\Delta:\Delta]$. 
Figure~\ref{f3}(a) shows the signed fraction of ${\rm sgn}\big({\rm Pf}(A)\big)$; 
the emergent ${\rm sgn}\big({\rm Pf}(A)\big)=1$ signals the gap closing associated with 
the condensation of the random flux state, predicted to occur near $\Delta\sim 0.96$\cite{Vojta2015}. 
Its Schmidt spectrum is exactly halved by ${\cal P}_D$ (see the inset), 
showing that the disordered eigenstates remain topologically ordered irrespective of 
${\rm sgn}\big({\rm Pf}(A)\big)=\pm 1$. 
\par
While the Pfaffian serves as a ${\mathbb Z}_2$ topological invariant in the quantum spin Hall effect~\cite{Kane2005,Fu2006}, 
the present case is more closely related to the Pfaffian invariant of the Kitaev chain, 
where it signals the U(1) symmetry breaking into the topological superconductor 
supporting unpaired Majorana edge modes~\cite{Kitaev2001}. 
This correspondence suggests that our framework naturally extends to general parton Hamiltonians 
and has similarity to the Kitaev chain. 

\subsection{Quasi-localized Majorana fermions} 
The Schmidt vectors encode the purely quantum many-body aspects of the QSL state. 
We divide the system into ${\cal A}={1,\cdots,k}$ and ${\cal B}={k+1,\cdots,N}$, and Schmidt-decompose the $r$th (=${\bm u,\bm n}$) eigenstate as
\begin{equation}
|{\Psi}_{r}\rangle
= \sum_{s=1}^{\chi}\lambda_s |\psi_{r,s}^{{\cal A}} \rangle |\phi_{r,s}^{{\cal B}}\rangle,
\label{eq:schmidt2}
\end{equation}
where $\chi$ is the number of nonzero Schmidt values. 
We define the subspace associated with the $r$th eigenstate as
${\cal V}_{k,r}^{{\cal A}}= {\rm span}\{|\psi_{r,s}^{{\cal A}}\rangle\}_{s=1}^\chi$.
Comparing two arbitrary states $r$ and $r'$, we find that if
${\cal V}_{k,r}^{({\cal A})} \cap {\cal V}_{k,r'}^{({\cal A})} =0$,
their Schmidt subspaces have no overlap in region ${\cal A}$.

Starting from $k=N$, we decrease $k$ until the overlap becomes nonzero. We define the ``$k$-local'' scale as the maximum size of ${\cal A}$ for which the two states share nontrivial local quantum states. If the states belong to different quantum-number sectors, $\bm u_r \ne \bm u_{r'}$, then $k$ is always smaller than the corresponding region size, and the states are trivially orthogonal. We therefore focus on pairs of states in the same $\bm u$ sector and plot $k_{rr'}$ in Fig.~\ref{f3}(b). 
For the Kitaev honeycomb model, states are ordered from left to right by the binary representation of the fermion occupation numbers ($n_1 n_2 \cdots n_{N/2 ; 2}$), while for the other models, they are ordered by energy.
The similarities between fermionic states differ markedly from those between the corresponding physical states after projection. For comparison, we perform the same analysis for the classical Ising, XX, Heisenberg (XXX), and XYZ spin chains, using all eigenstates for $N=10-13$. For the XYZ model, we set $J_x=1$, $J_y=1.53$, and $J_z=1.79$ in the Hamiltonian
$H_{{\rm XYZ}}=\sum_{\gamma=x,y,z}\sum_{i=1}^N J_\gamma S^\gamma_i S^\gamma_{i+1}$.
Compared with the Kitaev honeycomb model, the resulting density plots for these standard models exhibit relatively structureless patterns.

To quantify $k$-localness, we plot the distribution of $k_{rr'}/N$ over all pairs $(r,r')$ within 
the flux-free sector in Fig.~\ref{f3}(c). The Ising model exhibits a peak at $k=0$, with a distribution that decays exponentially with increasing $k$, reflecting the $2^{N-k}$ scaling of the number of $k$-local states. In contrast, the other quantum spin chains exhibit peaks at $k=N$, followed by exponential decay as $k$ decreases. This behavior may reflect the quantum correlations in these quantum-critical states, which arise from superpositions of similar states.
The $k$-localness of the Kitaev spin liquid, however, differs distinctly from both the Ising and other quantum spin states. The distribution vanishes for $k\gtrsim 0.9N$ and $k\le 0.5N$. The latter feature is attributed to the nonlocal Majorana fermionic wave function, which extends over more than half the system. Although the fermionic solutions of the Kitaev honeycomb model retain a nonzero component near $k=N$, their distribution differs from those of the other quantum states. This component vanishes upon projection onto the physical states.
Physically, this result indicates that the spin-liquid states do not share a common global basis set within the sector. This feature can be understood by analogy with a Rokhsar--Kivelson state\cite{Rokhsar1988}, an analytical realization of a ${\mathbb Z}_2$ spin liquid. Such a state is an equal-weight superposition of dimer coverings, and distinct basis states exhibit $k<N$-locality. These results establish $k$-localness as a fingerprint for identifying topological order in the quantum many-body framework. Indeed, the Kitaev honeycomb state exhibits a similar structure through the equal-weight superposition of $D_g$-operated states in Eq.~(\ref{eq:topowf}). 
\par
The spatial distribution of each Majorana fermion that constitutes the spin liquid ground state and excited states is also solved analytically (see Appendix \ref{sec:localmajorana}), 
and the maximum range that the fermions could spread is restricted to $\sim 3/4$ of the system size, 
which is consistent with the numerical observation of $k$-local in Fig.~\ref{f3}(c). 

\section{Parton theory} 
\subsection{Parton Hamiltonian}
To generalize our central findings to the QSL realized in other quantum spin models, 
we consider a mean-field parton Hamiltonian, which gives the basic description of the 
QSL obtained by the variational Monte Carlo studies. 
We start from the $S=1/2$ quantum Heisenberg model given as, 
${\cal H}= J\sum_{\langle i,j\rangle} {\bm S}_i \cdot {\bm S}_j$, 
and rewrite the spin operator using the Abrikosov fermion $f_{i\sigma}$ as 
\begin{align}
 {\bm S}_i=\frac{1}{2} f_{i\sigma}^\dagger \bm \sigma_{\alpha\beta}f_{i\beta}, \quad
 \sum_\sigma f_{i\sigma}^\dagger f_{i\sigma}=1, 
\end{align}
which gives the form, 
\begin{align}
{\cal H}_p^{0}= \frac{J}{4} \sum_{\langle i,j \rangle} f_{i\sigma}^\dagger
  f_{i\beta} f_{j\sigma}^\dagger f_{j\beta} \bm \sigma_{\alpha\beta}\cdot  \bm \sigma_{\alpha\beta}.
\label{eq:partonorg}
\end{align}
By introducing the mean-field as 
$\chi_{ij}=2\langle  f_{j\sigma}^\dagger f_{i\sigma} \rangle$, 
$\Delta_{ij}=\langle  -f_{i\uparrow} f_{j\downarrow} +f_{i\downarrow}  f_{j\uparrow} \rangle$, 
we find the parton Hamiltonian, 
\begin{align}
{\cal H}_{p}=&\sum_{\langle i,j\rangle} \Big(\sum_{\sigma=\uparrow,\downarrow}\chi_{ij}
 f_{i\sigma}^\dagger f_{j\sigma}  + \Delta_{ij} \big(f^\dagger_{i\uparrow} f^\dagger_{j\downarrow}  + h.c.\big)\Big) \notag \\
 & +  \sum_{i}  \Big( \sum_{\sigma=\uparrow,\downarrow} \mu f_{i\sigma}^\dagger f_{i\sigma}  + \zeta \big(f^\dagger_{i\uparrow} f^\dagger_{i\downarrow} + h.c. \big)\Big)
\label{eq:partonham}
\end{align}
After obtaining the eigenstate of this Hamiltonian, $|\Psi_{\rm MF}\rangle$, 
one needs to transform it to the spin space, which is done by 
the Gutzwiller projection, $P_G=\prod_i (n_{i\uparrow}-n_{i\downarrow})^2$, with 
$n_{i\sigma}=f_{i\sigma}^\dagger f_{i\sigma}$,  
and the resultant $|\Psi_{\rm SL}\rangle=P_G |\Psi_{\rm MF}\rangle$ is considered as 
a good ansatz for the spin liquid state realized in the original Heisenberg model. 
\subsection{Gauge operators} 
\label{sec:partongauge}
The Majorana fermions corresponding to the Abrikosov fermions are given as
\begin{align}
c_{i\sigma;1}=f_{i\sigma}+f^\dagger_{i\sigma}, \quad
c_{i\sigma;2}=(-i)(f_{i\sigma}-f^\dagger_{i\sigma}). 
\label{eq:majorana}
\end{align}
We introduce an operator 
\begin{align}
\hat D_i=c_{i\uparrow;1} c_{i\uparrow;2}c_{i\downarrow;1}c_{i\downarrow;2}. 
\end{align}
Since $\hat D_i$ commutes with ${\bm S}_i$ for all $i$, 
it is a conserved quantity for the spin Hamiltonian, or equivalently, ${\cal H}_p^{0}$ in Eq.(\ref{eq:partonorg}). 
The eigenstates of the spin Hamiltonian are indexed by $\{D_i=\pm 1\}_{i=1}^N$. 
When the eigenstate of ${\cal H}_p^{0}$ is obtained, it is put back into the spin space by the Gutzwiller projection, 
where only the $D_i=1,\,^\forall i$ component remains as the true ground state of the original Heisenberg model. 
However, the introduction of mean-field in Eq.(\ref{eq:partonham}) breaks this local conservation, as 
$\hat D_i$ no longer commutes with ${\cal H}_{p}$. 
Since $\mu$ in the mean-field Hamiltonian is set to keep the system half-filled, 
the expectation values of the following operators evaluated by their ground state $|\Psi_{f}^{(\rm MF)}\rangle$ satisfy, 
\begin{align}
\langle f^\dagger_{i \sigma} f_{i \sigma}\rangle=\frac{1}{2}, \quad
\langle f^\dagger_{i \uparrow} f_{i \downarrow}\rangle=\langle f^\dagger_{i \downarrow} f_{i \uparrow}\rangle=0. 
\end{align}
Here, we used $\langle f^\dagger_{i \sigma} f_{i \sigma}+ f_{i \sigma} f^\dagger_{i \sigma}\rangle=1$. 
Accordingly, $\langle c_j^a c_j^b \rangle=0$ is satisfied for all $a\ne b$, 
$a,b=(\alpha,\sigma)$ with $\alpha=1,2$, $\sigma=\uparrow,\downarrow$.  
Wick's theorem naturally gives, 
\begin{align}
\langle \hat D_i\rangle &= \langle c_{i\uparrow;1} c_{i\uparrow;2}  \rangle \langle c_{i\downarrow;1}c_{i\downarrow;2} \rangle
-\langle c_{i\uparrow;1}c_{i\uparrow;2} \rangle \langle c_{i\downarrow;1}c_{i\downarrow;2} \rangle
\notag \\
&\quad+\langle c_{i\uparrow;1}c_{i\downarrow;2} \rangle \langle c_{i\uparrow;2}c_{i\downarrow;1} \rangle
\notag \\
&=0, 
\end{align}
for the mean-field state that has equal probability of $D_i=\pm 1$. 
\par
The global projector is constructed as 
\begin{align}
\hat D=\prod_{i=1}^N \hat D_i, 
\end{align}
Unlike the local operator, $\hat D$ commutes with ${\cal H}_p$, so that 
$|\Psi_{f}^{(\rm MF)}\rangle$ is an eigenstate of $\hat D$, where only the $D=1$ value is allowed for fermions, 
because $D=-1$ does not survive after the Gutzwiller approximation. 
Accordingly, as has been done previously for the Kitaev-honeycomb case in Eqs.(\ref{eq:projection}) and (\ref{eq:topowf}), 
we are able to expand the post-projection wave function as, 
\begin{equation}
|\Psi_{f}^{(\rm MF)}\rangle= \frac{1}{\sqrt{2^{N-1}}} \sum_{\{D_i\},D=1} |\Psi_{f,\{D_i\}}\rangle, 
\end{equation}
where $|\Psi_{f,\{D_i\}}\rangle$ is the possible or approximate eigenstate of ${\cal H}_p^{0}$ before 
taking a mean-field, where each $D_i=\pm 1$ is well defined. 
\par
The nontrivial point for the general parton theory is that, unlike the Kitaev honeycomb model, 
$|\Psi_{f,\{D_i\}}\rangle$ is not necessarily written as the product of gauge and matter parts, 
which does not allow the discussions we expanded in Sec.~\ref{sec:topo-schmidt} and App.~\ref{app:schmidt}. 
However, because the global gauge operator in region ${\cal A}$, 
$\hat D_{\cal A}=\prod_{j=1}^{N_{\cal A}} \hat D_j$, 
also commutes the fermion parity operator in region ${\cal A}$ (see Eq.(\ref{eq:DA-parton})), 
similarly to Eq.(\ref{eq:DAeig}), 
the Schmidt state in ${\cal A}$ is an eigenstate of $\hat D_{\cal A}$. 
This fact allows us to expand the parallel discussions on the Schmidt decomposition we made for the Kitaev-honeycomb model. 
{\it The parton Hamiltonian therefore contains no explicit ${\mathbb Z}_2$ gauge variables.} 
\par
This $\hat D$ has the same form as Eq.(\ref{eq:z2eq}), 
except that the bond-gauge factor $\prod u_{j,k}$ is absent. 
When $\Delta_{ij}=\zeta=0$, the spin-up and down fermions remain decoupled, and ${\rm sgn}[{\rm Pf}(A)]=1$, 
so that the Pfaffian does not enter $\hat D$. 
After the Gutzwiller projection, this corresponds to a U(1) spin liquid for appropriate $\Delta_{ij}$, 
By contrast, finite pairing $\Delta_{ij}\neq 0$ introduces the factor ${\rm sgn}[{\rm Pf}(A)]=\pm 1$ into $\hat D$, 
yielding a ${\mathbb Z}_2$ spin liquid. 
The Pfaffian therefore encodes the breaking of the U(1) gauge symmetry 
(Appendix \ref{app:partonschmidt}). 

\begin{figure}[tbp]
   \centering
   \includegraphics[width=8.5cm]{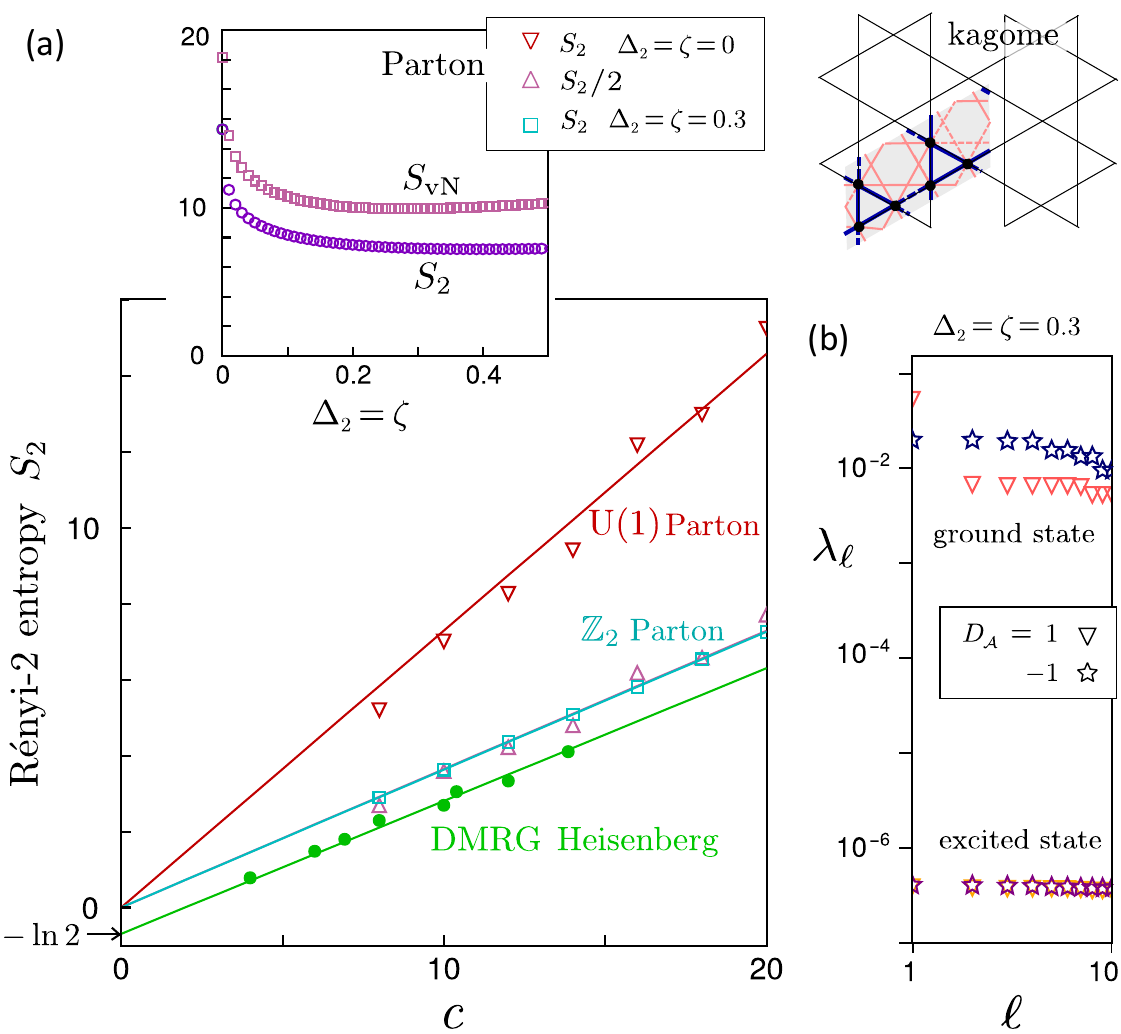}
    \caption{(a) R\'enyi-2 entropy $S_2$ of the kagome $S=1/2$ Heisenberg antiferromagnet, 
     using the parton theory in Eq.(\ref{eq:partonham})   
     and the DMRG ${\mathbb Z}_2$ spin liquid results taken from Ref.[\onlinecite{Schollwock2012}], 
     where we used the YC cylinders of $N_x=32$ and circumference $c=2N_y$, $N_y=4-10$. 
     Top right panel is the kagome lattice with the unit cell of the Parton state 
     whose parameters are $\chi_{ij}$ with nearest neighbor solid/broken bold lines are $\pm 1$, 
     next nearest solid/broken lines are $\pm 0.0186$\cite{Iqbal2011}. 
     We apply $\Delta_2=\zeta=0$ (U(1)) and $0.3$ (${\mathbb Z}_2$) 
     where the next nearest neighbor $\Delta_2$ also has $\pm$ signs. 
    Inset: $\Delta_2=\zeta$ dependence of the von Neumann entropy $S_{\text vN}$ and $S_2$ at $c=20$. 
    (b) Largest few $\lambda_\ell$ of the parton ${\mathbb Z}_2$ state ($\Delta_2=\zeta=0.3$) with $D_{\cal A}=\pm 1$, 
    in the center $N_{\cal A}=12$ subsystem of YC10 cylinder (see also Appendix \ref{app:partonnumerical}). 
    The line for the DMRG cylinder is taken from Ref.\cite{Schollwock2012}, and 
    the line for our ${\mathbb Z}_2$ is derived by the fitting of the data as $\alpha c + \gamma$, with $\gamma=0$. 
    The U(1) spin liquid has serious size effects and the fitting is performed by setting the zero intercept.
    }
\label{f4}
\end{figure}
%
\subsection{Kagome antiferromagnet} 
To show that the projection-induced topological order is not specific to the Kitaev honeycomb model, 
we analyze the U(1) and ${\mathbb Z}_2$ parton wave functions proposed for the kagome Heisenberg model~\cite{Iqbal2011} 
(Appendix \ref{app:partonnumerical}). 
As shown in the inset of Fig.~\ref{f4}, 
$\chi_{ij}$ between the nearest and next-nearest neighbors, 
second nearest neighbor $\Delta_{ij}$, 
and on-site $\zeta>0$ are considered, where both $\chi_{ij}$ and $\Delta_{ij}$ 
take plus and minus distributions. 
Figure~\ref{f4}(a) shows the R\'enyi-2 entropy $S_2$ on a YC cylinder as a function of the circumference $c$, 
the choice allowing direct comparison with the DMRG results for the ${\mathbb Z}_2$ spin liquid~\cite{Schollwock2012}. 
The ${\mathbb Z}_2$ parton state exhibits nearly half the $S_2$ of the U(1) state, 
which is understood in analogy with the Kitaev chain~\cite{Kitaev2001}; 
in the fermionic language, the superconducting pairing in the ${\mathbb Z}_2$ state halves 
the number of free particles, and hence their entanglement arising from their distribution is halved. 
\par
Remarkably, the unprojected ${\mathbb Z}_2$ parton state follows the almost the same linear scaling as the DMRG spin state, 
except for the constant $-\ln2$ shift: 
Here, notice that the linear line for DMRG is taken from Ref.\cite{Schollwock2012}, 
and the fitting curve of our data for the ${\mathbb Z}_2$ parton theory gives the slope with zero intercept.
Microscopically, the Schmidt spectrum separates into two nearly degenerate sectors 
with $D_{\cal A}=\pm 1$ (Fig.~\ref{f4}(b)), 
and the projection removes the $D_{\cal A}=-1$ sector, 
which we have proved analytically in Sec.~\ref{sec:partongauge} 
as the outcome of the commutativity of $\hat D_{\cal A}$ and ${\cal P}_D$. 
Similar behavior is found for the $J_1$-$J_2$ triangular-lattice antiferromagnet 
(see Fig.~\ref{fsparton}). 
As neither the Heisenberg models nor the parton Hamiltonian contains explicit bond-gauge variables, 
these results identify 
the post-projection collapse of the fermionic into spin state by the gauge constraint as 
a generic bulk signature of topological order. 
It is a bulk counterpart to the Wilson-loop characterization of topological order 
proposed by Levin and Wen~\cite{Levin2005,Levin2006}, establishing a bulk-boundary correspondence between 
the gauge constraint inside the wave function and the universal topological contribution to the EE. 
\section{Concluding remarks}
We have established an exact tensor-network representation of the full set of eigenstates of the Kitaev honeycomb model for sufficiently large systems to accurately capture their entanglement properties.
The complete Schmidt decomposition reveals that the Schmidt eigenvalues and corresponding eigenstates 
are reduced by half when a fermionic eigenstate undergoes Gutzwiller projection onto the physical spin Hilbert space. 
This reduction accounts for the universal $-\ln 2$ contribution to the entanglement entropy associated with topological order. 
\par
There are two remarks. 
First, since the topological contribution, at least for the gapped QSL, is taken as a key signature of a QSL, the unprojected Majorana fermion states are not the QSL state; the physical spin-liquid character emerges only after Gutzwiller projection. 
Second, this halving of the Schmidt spectrum occurs exactly for every eigenstate, 
indicating that the same topological entanglement contribution persists throughout the spectrum, including in excited states. 
\par
The mechanism underlying this halving is demonstrated by the exact analytical form of the Schmidt decomposition, 
using the global gauge operator $\hat D$, defined as the product of the local gauge operators $\hat D_i$ over all sites. 
Since $\hat D$ commutes with both the Majorana and spin Hamiltonians, 
its eigenvalue is factorized into those of subsystems $A$ and $B$, with $D=D_A D_B=1$ and $D_A,D_B=\pm1$. 
The Gutzwiller projection, which imposes the local constraints $D_i=1$ to reduce the enlarged Majorana Hilbert space 
to the physical spin Hilbert space, selects the $D_A=D_B=1$ sector. 
Importantly, this fact is rigorous, since the Gutzwiller projector commutes with both $\hat D_A$, $\hat D_B$. 
This constraint provides an analytical explanation for the reduction of the Schmidt spectrum by half and the resulting $-\ln 2$ contribution to the entanglement entropy.
\par
At first sight, this mechanism might appear to be a special feature of the exactly solvable Kitaev honeycomb model. 
However, we find that the same formulation of the gauge operator $\hat D$ applies  
to the Abrikosov-fermion parton mean-field theory, a standard framework for describing QSLs in the $S=1/2$ Heisenberg model. 
Guided by this observation, we investigate mean-field parton states of the kagome Heisenberg model using parameters obtained from previous variational Monte Carlo studies. 
The resulting Schmidt spectra exhibit near-degenerate sectors associated with $D_A=\pm1$, 
and the Gutzwiller projection is performed independently of the sectors, rigorously eliminating the $-1$ component. 
This finding suggests that selecting one of these sectors through Gutzwiller projection 
provides a microscopic origin for the topological entanglement contribution. 
However, an explicit Gutzwiller projection of the parton mean-field states for sufficiently large systems 
is not currently available. 
Indeed, previous VMC works show that the precise evaluation of the topological term, $\gamma$, 
with the Gutzwiller projection may suffer numerical difficulty, which can vary from $\gamma=0.584 \pm 0.089$\cite{Zhang2011} 
to $0.288 \pm 0.107$\cite{Zhang2011erratum}, and $\gamma=0.2\pm 0.04$\cite{Pei2013}
, even using the same scheme of evaluation. 
This fact highlights the utmost difficulty of even simply extracting the $-\ln 2$ term 
and performing the proper Gutzwiller projection to the quantum many-body level, which requires numerically exact 
schemes even with the limited system sizes. 
The application of the exact TN method to more general QSL states, in order to clarify the role 
of Gutzwiller projection to the general QSL needs further investigation. 

\par
Still, our findings provide a microscopic perspective on how gauge constraints shape the entanglement structure 
of QSL, extending previous loop-based approaches to a framework that directly connects gauge sectors with the Schmidt spectrum. 
More broadly, our results establish a basis for investigating how the gauge structure of parton representations determines the entanglement properties of physical spin states. 

\begin{acknowledgments}
We thank Ryui Kaneko for valuable comments. 
The work is supported by JSPS KAKENHI Grants No. JP26K00627, JP26H00635
JP21H05191, and JP21K03440 from the Ministry of Education, Science, Sports and Culture of Japan. H.S. is supported by the JSPS fellowship No. JP25KJ0762. 
\end{acknowledgments}
\appendix

\section{Details of exact tensor-network solutions}
\label{app:tn}
\subsection{Conserved quantities}
\label{sec:kitaev_rev}
In the Kitaev honeycomb spin Hamiltonian Eq.(\ref{eq:ham}), there are $N/2+1$ independent conserved quantities: 
the $N/2$ plaquettes $W_p=\prod_{i\in {\cal L}_p} \sigma_i^{\alpha_i}$ and the two loop operators $W_x=\prod_{i\in {\cal L}_x} \sigma_i^{\alpha_i}$, $W_y=\prod_{i\in {\cal L}_y} \sigma_i^{\alpha_i}$ where ${\cal L}_p$ 
is a loop around the plaquette $p$ and ${\cal L}_{x,y}$ are loops around the torus in the $x,y$ directions.
$\alpha_i=x,y,z$ is the bond index connnected to site $i$ but is not included in the loop. 
Since $\prod_{p} W_p=1$, the number of independent quantities decreases by one, 
and the Hilbert space has $2^{N/2+1}$ independent sectors of dimension $2^{N/2-1}$. 
\par
The conserved quantities $w=\{\{w_p\},w_x,w_y\}=$ 1 or $-1$ are the product of $u_{jk}$ on the loop. 
We note that since $|\tilde{\Psi}_{u,n} \rangle$ is an eigenstate of $\hat D=\prod_{i=1}^N \hat{D}_i$, 
$\bm u,\bm n$ needs to satisfy $\hat{D}=1$ to obtain a physical state\cite{Loss2011,Yao2009}, 
where $D$ is explicitly written as \cite{Loss2011}
\begin{equation}
D= \prod_{i=1}^N D_i=(-1)^{N_x}\det Q_r \prod_{l=1}^{N/2}(1-2n_l)\prod_{\langle j,k\rangle}u_{jk},
\label{eq:prod_D}
\end{equation}
where $Q_r=\sqrt{2} [{\rm Re}Q_1, {\rm Im}Q_1, \cdots, {\rm Re}Q_{N/2}, {\rm Im}Q_{N/2}]$ is an orthogonal matrix, 
whose columns are the real and imaginary parts of half the columns of $Q$ used to diagonalize the free Majorana, 
$Q=[Q_1,\cdots,Q_N]$. 
\par
By converting $\det Q_r=(-)^{N/2} \mathrm{sgn}(\mathrm{Pf}(A))$, 
we obtain the explicit form Eq.(\ref{eq:z2eq}). 
Here, we briefly explain its derivation. 
Because the antisymmetric matrix $A$ is block-diagonalized as 
$Q_r^{t}A Q_r$, with $\mp \epsilon_\ell$ 
forming an off-diagonal element of each $\ell=1,\cdots, N/2$ block matrices, 
\begin{align}
\mathrm{Pf}(Q_r^{t}A Q_r)=(-)^{N/2} \prod_{\ell=1}^{N/2}\epsilon_\ell=(-)^{N/2} \sqrt{\det A}, 
\end{align}
where $\det(iA)=(\prod_{\ell=1}^{N/2}\epsilon_\ell)(\prod_{\ell=1}^{N/2}-\epsilon_\ell)$. 
Using the formula $\mathrm{Pf}(Q_r^{t}A Q_r)=\det(Q_r)\mathrm{Pf}(A)$, 
we obtain 
\begin{align}
\det(Q_r)=(-)^{N/2}\sqrt{\det A}/\mathrm{Pf}(A)=(-)^{N/2}\mathrm{sgn}(\mathrm{Pf}(A)). 
\end{align}
\subsection{Distribution of singular values}
\label{sec:dist}
\begin{figure}
    \centering
    \includegraphics[width=7.5cm]{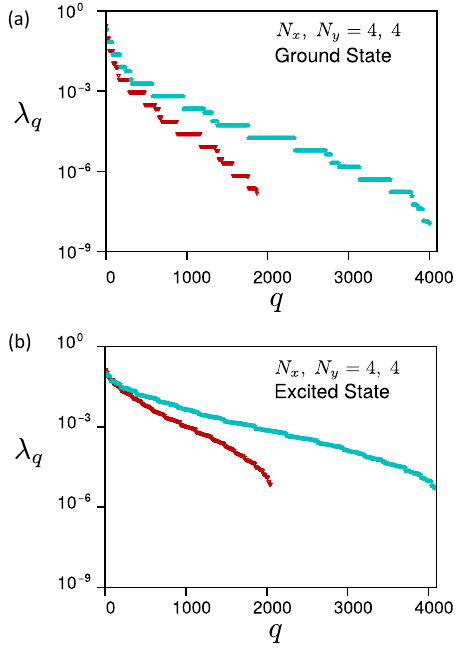}
    \caption{
        Schmidt values of the eigenstates of the Kitaev honeycomb model with $J_x=J_y=J_z=1$ obtained from 
         our exact TN solutions 
        where $N=2N_x\times N_y=2\times 4\times 4$ and we divide the system in half. 
        For the fermionic eigenstate, $|\tilde{\Psi}^M_{\bm u,\bm n}\rangle$, we find 4096 nonzero Schmidt values 
        (blue symbols), which reduces to 2048 for the physical eigenstate $|\Psi_{\bm u,\bm n}\rangle$ (red symbols). 
        We plot those of (a) the ground state and (b) the excited state sector with $\bm u,\bm n$ randomly chosen. 
    }
    \label{fs2}
\end{figure}

We discussed in the main text that for all the TN exact solutions, 
the bond dimension $\chi_m$ which is the number of nonzero Schmidt coefficients 
is the same regardless of whether it is the ground state or the excited state 
(for any given quantum numbers $\bm u,\bm n$). 
Figure~\ref{fs2} shows the distribution of nonzero singular values $\lambda_q$ when we Schmidt-decompose the state 
by the bipartition 
to ${\cal A} = \{1,\cdots,m\}$ and ${\cal B}=\{m+1,\cdots,N\}$ where we set $m=N/2$. 
Two distinct features are observed. 
For all values of $\lambda_q$, the degeneracy is twice as large for the 
fermionic eigenstate, $|\tilde{\Psi}^M_{\bm u,\bm n}\rangle$, while the corresponding $\lambda_q$'s 
of the physical eigenstate $|\Psi_{\bm u,\bm n}\rangle$ is reduced, $\lambda_q\rightarrow \sqrt{2}\lambda_q$. 
Namely, by scaling the horizontal and vertical axes by $\times 2$ and $/\sqrt{2}$, the two data lines exactly collapse. 
\par
The second feature is that although the maximum number of Schmidt values $\chi_m$ is the same 
between the ground and excited states, the values of $\lambda_q$ are more widely spread for the latter, 
which yields a volume-law entanglement. 
\subsection{Volume law of the excited states}
\label{sec:volumelaw}
\begin{figure}
    \centering
    \includegraphics[width=8.5cm]{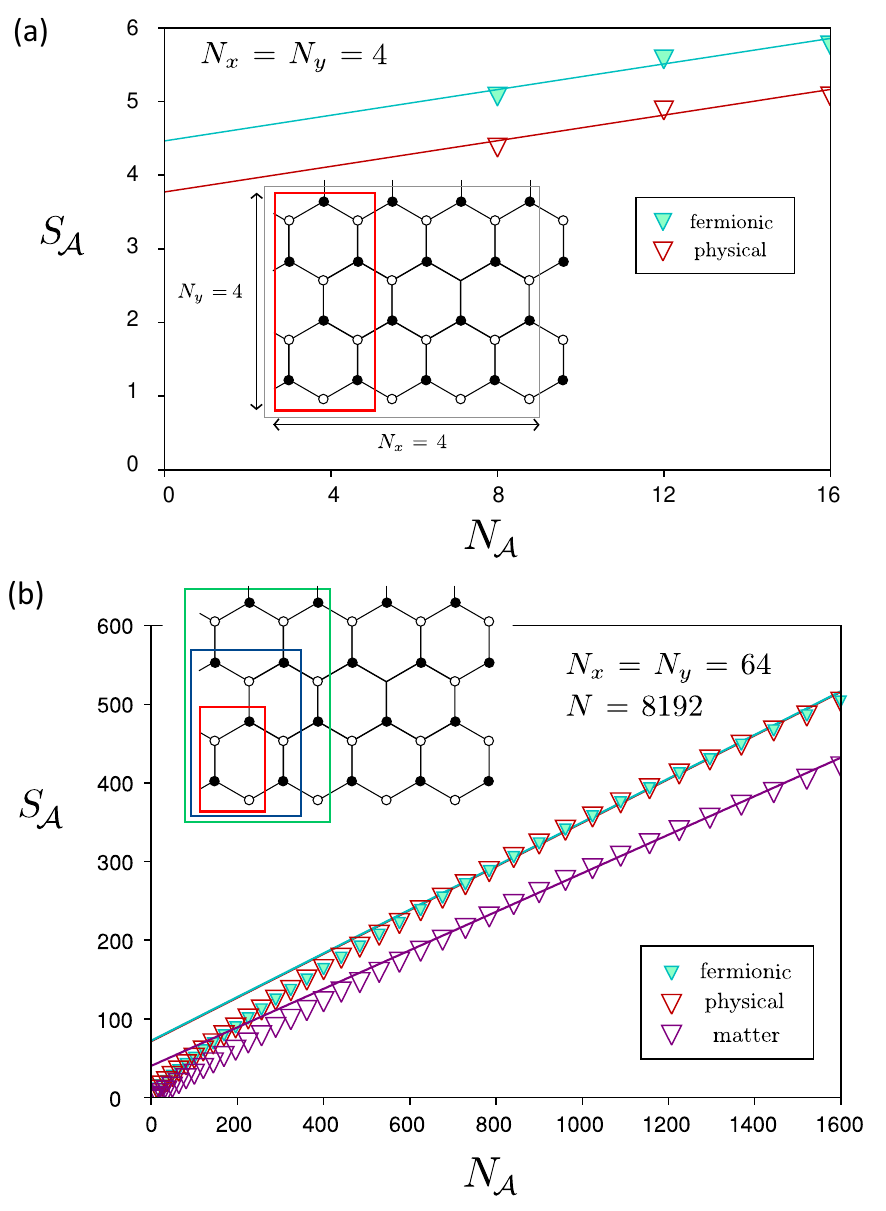}
    \caption{
     Entanglement entropy $S_{\cal A}$ of the excited states of the Kitaev honeycomb model with $J_x=J_y=J_z=1$. 
      (a) $N=2 N_x \times N_y= 2\times 4\times 4=32$ with $N_{\cal A}$ sites 
      in region ${\cal A}$ with fixed $N_y=4$. 
      (b) $N=2 N_x \times N_y= 2\times 64\times 64=8192$ with $N_{\cal A}=n^2$ taken as in the inset. 
       Contributions from matter fermion, $S_{\cal A}^m=\sum_\ell \lambda^2_\ell \ln \lambda_\ell^2$, 
       from the Majorana fermions, $S_{\cal A}^f= L\ln 2 + S_{\cal A}^m$, 
       and from the physical state after the Gutzwiller projection, 
       $S_{\cal A}^p=(L-1)\ln 2 + S_A^m$, are separately shown. Solid lines are the linear fits that show the volume law.
       Calculation is done by using exact TN representation for (a) and by analytical formula for (b). 
    }
    \label{fsvol}
\end{figure}

In Fig.~\ref{f2}(a) in the main text, we showed the entanglement entropy for $N=32$ 
of the MPS-MPO solution, where we argued that the volume law behavior is observed for the 
excited state solutions in contrast to the ground state, 
although the bond dimensions are the same.  
To further support our claim, we show in Fig.~\ref{fsvol}(a) 
the entanglement entropy of the same $N=2N_x\times N_y=2 \times 4\times 4$ cluster 
as a function of $N_{\cal A}=2N_x^{\cal A}N_y^{\cal A}$ (the number of sites included in ${\cal A}$), 
where we set $N_y^{\cal A}=4$ constant and vary $N_x^{\cal A}$. 
The results do not contradict the volume law. 
\par 
For clarification, we show in Fig.~\ref{fsvol}(b) the case of $N_x=N_y=64$ that amounts to $N=8192$, 
with the region ${\cal A}$ taken as in the inset, keeping the aspect ratio to 1 in the $x$ and $y$ directions; 
$N_{\cal A}=n^2$ with the boundary length given as $L=3n$. 
We compare the case of the entanglement entropy of the matter fermions, 
$S_{\cal A}^m=\sum_\ell \lambda^2_\ell \ln \lambda_\ell^2$, 
that of the Majorana fermions, 
$S_{\cal A}^f=L\ln 2 + S_{\cal A}^m$, 
and the physical ones after the Gutzwiller projection, 
$S_{\cal A}^p=(L-1)\ln 2 + S_{\cal A}^m$. 
The difference, $S_{\cal A}^p-S_{\cal A}^f= -\ln 2$, is almost invisible for the scale of the figure, 
while it is present and supports the topological order. 
With this result, one finds that the volume law is safely distinguished 
at $N_{\cal A}\gtrsim 600$, and is present for larger $N$. 
We also notice that the matter fermion part, $S_{\cal A}^m$ is mainly responsible for 
the volume-law behavior of the excited states of the Kitaev honeycomb model. 

\section{Gutzwiller projection}
\subsection{Gauge constraint and post-projection state. }
\label{app:gauge-constraint}
Let us illustrate the nature of the post-projection state. 
The fermionic state is an eigenstate of $\hat D=\prod_{i=1}^N \hat D_i$ 
but is not the eigenstate of $\hat D_i$. 
The operator ${\cal P}_D$ in Eq.(\ref{eq:pd}) projects it to the simultaneous eigenstates of all $\hat D_i$'s 
and is expanded as
\begin{align}
{\cal P}_D =\prod_{j=1}^N \frac{1+\hat D_j}{2}  =\frac{1}{2^N} \sum_g \hat D_g ,  \quad \hat D_g=\prod_{j\in g} \hat D_j, 
\label{eq:projection}
\end{align}
where $g$ runs over all the subsets of lattice sites, 
chosen from among all the lattice sites ${\mathcal L}$. 
Because the condition $\hat D|\tilde\Psi^{M}_{\bm u,\bm n} \rangle =|\tilde\Psi^{M}_{\bm u,\bm n} \rangle$, 
is rewritten as 
$\hat D_g|\tilde\Psi^{M}_{\bm u,\bm n} \rangle = \hat D_{{\mathcal L}-g} |\tilde\Psi^{M}_{\bm u,\bm n} \rangle$, 
among the $2^N$ terms in $\sum_g$, only half are independent. 
The normalization factor of the post-projected state needs to be multiplied by $1/\sqrt{2}$ to reduce this redundancy: 
\begin{align}
|\Psi_{\bm u,\bm n} \rangle
& = \frac{1}{\sqrt{2^{N+1}}}\sum_{g} \hat D_g |\tilde \Psi^{M}_{\bm u,\bm n} \rangle, 
\label{eq:topowf}
\end{align}
which ends up with the equal-weight superposition of $2^{N-1}$ independent states 
gauge-transformed by different $\hat D_g$'s \cite{Yao2010}. 
These discussions rephrase those in Refs.[\onlinecite{Yao2009,Loss2011,Yao2010}]. 
\par
To prepare for the next subsection, we now divide the system into ${\cal A}$ and ${\cal B}$. 
Here, the projection operator, ${\cal P}_D$, 
is naturally divided into those operating on the two as ${\cal P}_D={\cal P}_D^{\cal A}{\cal P}_D^{\cal B}$, with 
\begin{align}
{\cal P}_D^{\cal A} =\prod_{j=1}^{N_{\cal A}} 
\frac{1+\hat D_j}{2}  =\frac{1}{2^{N_{\cal A}}} \sum_{g\in {\cal A}} \hat D_g ,  
\quad \hat D_g=\prod_{j\in g} \hat D_j. 
\end{align}
Using this factorization, we apply ${\cal P}_D$ to Eq.(\ref{eq:schmidt}) and obtain 
the explicit form of the Gutzwiller projected eigenstates of the original Hamiltonian, Eq.(\ref{eq:Schmidtall}). 
By further expanding ${\cal P}_D^{\cal A}$ and ${\cal P}_D^{\cal B}$ in terms of $\hat D_i$, 
we obtain the following form, 
\begin{align}
|\Psi_{\bm u,\bm v}\rangle & =\sum_g \sum_{\{w_{\cal A}\},D_{\cal A}=1} \sum_{\ell}  \frac{\lambda_\ell}{\sqrt{2^{N+L+1}}}
   D_g |\psi^{\cal A}_{\ell}\rangle |\psi^{\cal B}_{\ell}\rangle.
\end{align}

\subsection{How the Schmidt states are reduced by half. } 
\label{app:schmidt}
Here, we explain the details of the derivation of Eq.(\ref{eq:schmidt}) in the main text and related results. 
Let us divide the system into subsystems ${\cal A}$ and ${\cal B}$ by drawing a closed loop as shown in Fig.~\ref{f1}(e) in the main text. 
Suppose that this closed loop cuts $2L$ bonds on the lattice, which are indexed as $1,2,\cdots, 2n-1,2n,\cdots 2L$, 
where we relabel the bond operator $\hat u_j$ as $\hat u_{2n-1}$ and $\hat u_{2n}$ 
labeled by $\alpha,\beta=x,y,z$, 
and the corresponding fermionic operator $\hat b^{\alpha}_{{\cal R},2n-1}$ and 
$\hat b^\beta_{{\cal R},2n}$ that belong to that bond on the $\bm R={\cal A, B}$ side. 
Next, we rearrange the bond operator on the $n$-th hexagonal unit 
and define a set of states $|w_{{\cal A},n}, w_{{\cal B},n}\rangle$ as 
\begin{align}
& \hat f_{1n}=\frac{1}{2}(\hat b_{{\cal A},2n-1}^\alpha + i\hat b_{{\cal A},2n}^\beta), \quad
  \hat f_{2n}=\frac{1}{2}(\hat b_{{\cal B},2n-1}^\alpha + i\hat b_{{\cal B},2n}^\beta), 
\notag\\
&\bm w_n= |w_{{\cal A},n}, w_{{\cal B},n}\rangle=\left\{
\begin{array}{l}
|1,-1\rangle = \hat f^\dagger_{1n} |0\rangle \\
|-1,1\rangle = -i\hat f^\dagger_{2n} |0\rangle \\
|1,1\rangle = \hat f^\dagger_{1n}\hat f^\dagger_{2n} |0\rangle \\
|-1,-1\rangle = i |0\rangle \\
\end{array}
\right.
\end{align}
From this definition, the original gauge states 
$\bm u_n =|u_{2n-1},u_{2n}\rangle= (|-1,-1\rangle , |1,1\rangle, |1,-1\rangle, |-1,1\rangle )^T$ 
are related to $\bm w_n$ as 
\begin{align}
\bm u_n =\frac{1}{\sqrt{2}} \left( \begin{array}{rrrr}
1 & -1  \\
1 &  1  \\
& & 1 & -1 \\
& & 1 & 1 
\end{array}
\right)
\bm w_n. 
\label{eq:utow}
\end{align}
Therefore, for a given $\bm u_n$, $|w_{{\cal A},n}, w_{{\cal B},n}\rangle$ has $\mathbb Z_2$ degrees 
of freedom which we label $w_n=\pm 1$; when $u_{2n-1}=u_{2n}$ it corresponds to the choice $|1,-1\rangle$ and $|-1,1\rangle$, 
and when $u_{2n-1}=-u_{2n}$, $|1,1\rangle$ and $|-1,-1\rangle$. 
Because a set of bond gauges $\bm u$ forms a product state, 
with this local gauge transformation around the loop, it can be separated into 
those inside region ${\cal R}$ denoted as $|\bm u^{\cal R}\rangle$ and those on the boundaries 
which are rearranged into $|w_{{\cal A},n}, w_{{\cal B},n}\rangle$. 
\par
The separation of the bond-gauges into region ${\cal A}$ and ${\cal B}$ so far discussed 
are based on Ref.[\onlinecite{Yao2010}] by Yao and Qi. 
In their framework, they applied the form Eq.(\ref{eq:topowf}) to describe the density matrix, 
$\rho_{\cal A} = {\rm Tr}_B(|\Psi_{\bm u,\bm n} \rangle\langle \Psi_{\bm u,\bm n} |)$, 
where they operated $\hat D_g = \hat X_{g{\cal A}} \hat Y_{g{\cal A}}\hat X_{g{\cal B}}\hat Y_{g{\cal B}}$, 
separately on the subsystems, each on bond gauge and matter parts. 
In this form (Eq.(A7) of the Supplemental \cite{Yao2010}), the fermion parity in the matter state 
is constrained by the partial Gutzwiller projection $\hat Y_{g{\cal A}}$ and $\hat Y_{g{\cal B}}$. 
Among the matter states, the ones that survive depend explicitly on the boundary gauge configuration, 
which then appears in their form of $\rho_{\cal A}$ as off-diagonal element of the projection operators,  
$\hat X_{g{\cal A}} |u_{\cal A},w\rangle\langle u_{\cal A},w|\hat X_{g{\cal A}'}^\dagger$. 
In such a case, the Schmidt decomposition appears no longer to be carried out independently of the gauge sector, 
making the summation over $g$ and $g'$ (gauge transformation in Eq.(\ref{eq:topowf}) that appear inside $\rho_{\cal A}$) highly nontrivial. 
Instead, Yao and Qi evaluated the integrated value ${\rm Tr}_{\cal A}[ (\rho_{\cal A})^n]$, applied the Replica trick, 
$S=-\frac{\partial}{\partial n} {\rm Tr}_{\cal A}[ (\rho_{\cal A})^n]|_{n=1}$, 
and obtained $S=(L-1)\ln 2+\alpha L$. 
\par
Our approach proceeds in the opposite order. 
{\it Before applying the Gutzwiller projector}, we first perform the exact Schmidt decomposition of the fermion wave function 
for both the matter and bond gauge parts (Eq.(\ref{eq:schmidt}) in the main text), 
\begin{widetext}
\begin{align}
|\tilde \Psi^M_{\bm u,\bm v}\rangle&= |\bm u\rangle|M_{\bm u}\rangle 
= \Big( \sum_{\ell=1}^\chi \lambda_\ell |M^{A}_{\bm u \ell}\rangle |M^{B}_{\bm u \ell}\rangle \Big)
|\bm u^{A}\rangle |\bm u^{B}\rangle \sum_{\{w_n=\pm 1\}} \prod_{n=1}^{L} |w_{A,n}, w_{B,n}\rangle
\notag \\
&=\sum_{\ell=1}^\chi \sum_{\{w_n=\pm 1\}} (-)^\gamma
\frac{\lambda_\ell}{\sqrt{2^{L}}}  \Big(  |M^{{\cal A}}_{\bm u \ell}\rangle |\bm u^{{\cal A}}\rangle |w^{{\cal A}}\rangle \Big)
\Big( |M^{{\cal B}}_{\bm u \ell}\rangle |\bm u^{{\cal B}}\rangle |w^{{\cal B}}\rangle \Big), 
\label{eq:schmidt2}
\end{align}
\end{widetext}
where the sign $(-)^\gamma$ comes from the combination of either of the two choices of 
$|w_{{\cal A},n}, w_{{\cal B},n}\rangle$ values in Eq.(\ref{eq:utow}). 
The relevant summation $\sum_{w_n=\pm 1}$ means that for each bond on a cut, there are two equivalent 
contributions from the choices of rearranged $\mathbb Z_2$ bond gauges, 
that generate $2^L$ different terms. 
This contribution explains the highly degenerate Schmidt values found numerically in Fig.~\ref{f2}(d) in the main text. 
\par
Since a set of $\{u_j\}$ are fixed values, the boundary gauges on the $2L$ bonds 
should naturally satisfy  $\prod_{n}w_{\cal A}= \eta \prod_{n}w_{\cal B}=\pm 1$ (where $\eta$ is either $+1$ or $-1$) 
under the $D=1$ constraint. The two choices of the boundary gauge have one-to-one correspondence 
with the choice of $D_{\cal A}=\pm 1$. 
\par
In Eq.(\ref{eq:schmidt2}), Schmidt decomposing $|M_{\bm u}\rangle$ for a fixed bond gauge 
(classical variables) is trivial itself. 
However, as we clarify in the following equations, $|M^{{\cal R}}_{\bm u \ell}\rangle$ 
are classified by the fermion parity in region ${\cal R=A,B}$, 
which has one-to-one correspondence with the boundary gauge sectors, which may not be trivial. 
This finding is a key to performing a full Schmidt decomposition of the physical state after the Gutzwiller projection 
(see Eq.(\ref{eq:Schmidtall})). 
\par
We now consider the global gauge operator, factorized into two parts, $\hat D=\hat D_{\cal A} \cdot \hat D_{\cal B}$, where 
$\hat D_{\cal R}=\prod_{j\in {\cal R}} \hat D_j$. 
The one operating on ${\cal A}$ can be expressed using a fermionic operator as
\begin{align}
\hat D_{\cal A}&=\prod_{j\in {\cal A}} \hat D_j=\prod_{j\in {\cal A}} \hat b_j^x\hat b_j^y\hat b_j^z \hat c_j 
\notag\\
&= (-)^m \prod_{u_{ij}\in {\cal A}} \hat u_{ij} \prod_{n=1}^L \hat w_{{\cal A},n} \prod_{l\in {\cal A}}\hat c_l, 
\end{align}
with the sign $(-)^m$ properly chosen. 
Since the matter Schmidt states, $|M^{\cal R}_{\bm u \ell}\rangle$, on region ${\cal A}$ and ${\cal B}$ are both 
free Majorana states, 
they are the eigenstates of $\prod_{l \in {\cal R}} \hat c_l$ with eigenvalues $\pm 1$; 
the operator $Q$ that transforms Majorana to fermions has 
$\prod_\ell c_\ell={\rm det}(Q)\prod_\ell (1-2n'_{\ell})$ with $n'_{\ell}$ the 
fermion number operator (see Ref.[\onlinecite{Loss2011}] Appendix A for details). 
Therefore, one can straightforwardly show that, 
\begin{equation}
\hat D_{\cal A} |M^{{\cal A}}_{\bm u \ell}\rangle |\bm u^{{\cal A}}\rangle |w^{{\cal A}} \rangle
= \pm |M^{{\cal A}}_{\bm u \ell}\rangle |\bm u^{{\cal A}}\rangle |w^{{\cal A}}\rangle, 
\label{eq:DAeig}
\end{equation}
where $\pm 1$ eigenvalues come from the two choices of boundary gauge states $|w^{\cal A}\rangle$, 
as $|u^{\cal A}\rangle$ is unique about $\hat u_{ij}$. 
\par
With these informations, one can say that $|\tilde \Psi^M_{\bm u,\bm v}\rangle$ is the $D=1$ eigenstate 
and its Schmidt state $\{|\psi^{\cal A}_\ell\rangle\}$ includes two eigenstates of $D_{\cal A}=\pm 1$. 
However, since ${\cal P}_D$ enforces $D_{\cal A}=1$, only half of them remain, 
which explains that the number of Schmidt states is reduced by half for each $\lambda_\ell$ and 
accordingly, $\lambda_\ell$ is scaled by $\sqrt{2}$. 
Using Eq.(\ref{eq:DAeig}), one can separately operate the Gutzwiller projection to ${\cal A}$ and ${\cal B}$ parts 
of the Schmidt form in $|\tilde \Psi^M_{\bm u,\bm v}\rangle$,  
and $|\Psi_{\bm u,\bm v}\rangle$ of the spin Hamiltonian is also Schmidt decomposed as 
in Eq.(\ref{eq:Schmidtall}) of Sec.~\ref{sec:topo}. 
%
\section{Spatial range that the Majorana fermion occupies}
\label{sec:localmajorana}
We want to obtain a quantum number of subsystem $\cal A$ that characterizes this subspace. 
The value of $\{\bm u\} \in {\cal A}$ is local and trivial, 
whereas $a_\ell/a_\ell^\dagger$ acts on all the sites, 
so that the number of localized Majorana fermions needs to be calculated. 
\par
Suppose the free Majorana state before projection, $|\tilde{\Psi}^M_{r}\rangle \equiv |\tilde{\Psi}^M_{u,n}\rangle$, 
has an $\ell$ th Majorana, $n_\ell=1$ (or not, $n_\ell=0$), 
which is read off as $\hat{a}_\ell^\dag |\tilde{\Psi}^M_{r}\rangle=0$ 
(or $\hat{a}_\ell |\tilde{\Psi}^M_{r}\rangle=0$). 
For a given set of values, $\{n_\ell\}_{\ell=1}^{N/2}$, 
one can define a vector $\hat{\bm{a}} = \{\hat{a}_l^{\times}\}_{l=1\cdots,N/2}$, where $\times=\dag$ for $n_l=1$ and $\times$ means nothing otherwise. 
Then, the condition that the state needs to satisfy to have these Majorana numbers is given as 
a set of linear equations, $\hat{\bm{a}}|\tilde{\Psi}^M_{r}\rangle=\bm 0$. 
By substituting $\hat{a}_\ell=1/\sqrt{2}\sum_{j=1}^N Q_{j\ell} c_j$ with $\bm{c}=\{c_j\}_{j=1\cdots,N}$, we find
\begin{align}
    \tilde{Q} \bm{c}|\tilde{\Psi}^M_{r}\rangle = 0,
\end{align}
where the $\tilde{Q}=\{\tilde{Q}_{\ell j}\}$ is the $N/2\times N$ matrix 
that satisfies $\tilde{Q}_{\ell j}=Q^\ast_{j\ell}$ (or $Q_{j\ell}$) for $n_\ell=1$ (or 0). 
This matrix is rewritten as 
\begin{align}
\tilde{Q} = UL, 
\end{align}
using the unitary matrix $U$ and lower triangular matrix $L$. 
One then finds that a new set of Majorana fermions, $\hat{\bm{a}}'=\{\hat{a}_l'\}_{l=1\cdots,N/2}$, 
transformed from $\hat{\bm{a}} = U \hat{\bm{a}}'$, 
and because of the lower triangular form of $L$, $\bm a'$ acts only on the subsystem ranging within $[1:N/2+\ell]$. 
It means that one can specify which Majorana, $\ell$, is localized in region ${\cal A}$ ($1:k$) 
by judging that if $k \ge N/2+\ell$, it satisfies $\hat{a}_l'^\dag\hat{a}_l'=0$, and otherwise nonlocalized. 
We could list the localized Majoranas and fully characterize ${\cal V}_{k,r}^{\cal A}$ accordingly. 
One important conclusion here is that all the Majoranas are 
spread over more than half the system size, regardless of the shape or size of the cluster.

\section{Details of Parton theory}
\subsection{Schmidt decomposition}
\label{app:partonschmidt}
We first Fourier transform Eq.(\ref{eq:partonham}) as, 
$f_{\bm k s \sigma}=N_c^{-1/2} \sum_{l=1}^{N_c} e^{i\bm k \bm r_{l,s}} f_{l s \sigma}$ 
where we redefined  $f_{l s \sigma}$ as the Abrikosov fermion in the $l (=1,\cdots,N_c)$ th unit cell on $s(=1,\cdots,n_s)$ th site 
($j\in (l,s)$). 
For the kagome lattice, $N_c=N/6$ and $n_s=6$, shown in Fig.~\ref{fsparton}(a). 
The Hamiltonian is transformed as 
\begin{align} 
{\cal H}_{p}=&\sum_{\bm k,\sigma} \sum_{s,s'} f_{\bm k s \sigma}^\dagger W_{s,s'}(\bm k) f_{\bm k s' \sigma} 
\notag\\
&\quad + \sum_{\bm k}  \sum_{s,s'} (f^\dagger_{\bm k s \uparrow} D_{s,s'}(\bm k) f^\dagger_{-\bm k s'\downarrow} + h.c.). 
\end{align}
When $\Delta_{ij}=\zeta=0$, we find $D_{s,s'}=0$ so that we only need to diagonalize the $n_s \times n_s$ matrix $W$ 
with $Q^\sigma, \sigma=\uparrow,\downarrow$, 
as $Q^{\sigma^\dagger} W Q^\sigma ={\rm diag} (\epsilon_\ell(\bm k))_{\ell=1,\cdots n_s}$, 
with the corresponding fermions given as, $\tilde f_{\bm k \ell \sigma}= \sum_s Q_{\ell s}^{\sigma \dagger} f_{\bm k s \sigma}$. 
\par
Whereas when $\Delta_{ij}\ne 0$, the Hamiltonian has a Bogoliubov-de-Gennes form, 
\begin{align}
{\cal H}_{p}&=\sum_{\bm k>0} \bm \Psi_{\bm k}^\dagger H (\bm k) \bm \Psi_{\bm k}, 
\notag\\
& H (\bm k)=  \left(
\begin{array}{rr}
W (\bm k) & D (\bm k) \\
D^{t*}(\bm k) & -W(\bm k)
\end{array}
\right) ,
\notag\\
& \bm\Psi_{\bm k}^\dagger =( f^\dagger_{\bm k,1,\uparrow},\cdots,f^\dagger_{\bm k,n_s,\uparrow},\cdots, 
f_{-\bm k,1,\downarrow},\cdots,f_{-\bm k,n_s,\downarrow}), 
\label{eq:BdG}
\end{align}
and by diagonalizing the $2n_s \times 2n_s$ matrix $H$ as 
$Q^\dagger H Q= {\rm diag}(\{-w_{\bm k,\ell}\}, \{w_{\bm k,\ell}\})_{\ell=1,\cdots,n_s}$, 
we obtain a Bogoliubov fermion $(b_{\bm k,\ell}, b_{-\bm k \ell}^\dagger) = Q^\dagger \Psi_{\bm k}$, 
and the corresponding superconducting energy band $\pm w_{\bm k,\ell}$. 
Here, we set the chemical potential to have the system half-filled, 
which is inside the band gap. 
\par
We now prepare a correlation matrix $C$ for the Majorana fermions in Eq.(\ref{eq:majorana}),
 $\{c_{l s \sigma;\alpha}\}$ with ($\alpha=1,2$) as 
\begin{widetext}
\begin{align}
& C_{(l s \sigma; \alpha, l' s'\sigma'; \alpha')}
  =\langle c_{l s \sigma ;\alpha}c_{l' s'\sigma' ;\alpha'} \rangle, 
\\
& \quad \quad \langle c_{l s \sigma ; 1} c_{l' s' \sigma' ;1} \rangle 
= \langle f_{l s \sigma}^\dagger f_{l' s' \sigma'}
       +  f_{l s \sigma} f_{l' s' \sigma'}^\dagger 
       +  f_{l s \sigma}^\dagger f_{l' s' \sigma'}^\dagger 
       +  f_{l s \sigma} f_{l' s' \sigma'} \rangle,
\notag \\
& \quad \quad \langle c_{l s \sigma ; 2} c_{l' s' \sigma' ;2} \rangle 
= \langle f_{l s \sigma}^\dagger f_{l' s' \sigma'}
       +  f_{l s \sigma} f_{l' s' \sigma'}^\dagger 
       -  f_{l s \sigma}^\dagger f_{l' s' \sigma'}^\dagger 
       -  f_{l s \sigma} f_{l' s' \sigma'} \rangle,
\notag \\
& \quad\quad \langle c_{l s \sigma ;1} c_{l' s' \sigma' ;2} \rangle
= (-i) \langle f_{l s \sigma}^\dagger f_{l' s'  \sigma'}
       -  f_{l s \sigma} f_{l's'  \sigma'}^\dagger 
       -  f_{l s \sigma}^\dagger f_{l' s' \sigma'}^\dagger 
       +  f_{l s \sigma} f_{l' s' \sigma'} \rangle
\notag\\
& \quad\quad \langle c_{l s  \sigma;2} c_{l' s'  \sigma';1} \rangle
= i \langle f_{l s  \sigma}^\dagger f_{l' s'  \sigma'}
       - f_{l s \sigma} f_{l's' \sigma'}^\dagger 
       +  f_{l s \sigma}^\dagger f_{l' s' \sigma'}^\dagger 
       -  f_{l s \sigma} f_{l' s' \sigma'} \rangle, 
\end{align}
\end{widetext}
where $\langle f_{l s \sigma}^\dagger f_{l' s' \sigma'}\rangle$ is obtained by transforming its representation to Bogoliubov fermions. 
\par
When dividing the system into subsystems ${\cal A}$ and ${\cal B}$, the Schmidt values are given as a submatrix of $C$, 
whose column and row indices belong to region {\cal A}: 
$C_{\cal A}=C_{(m,m')\in {\cal A}}\equiv \delta_{m m'}+i(\Gamma_{\cal A})_{m m'}$, with $m=(l s \sigma \alpha)$
Diagonalizing $C_{\cal A}-1=i\Gamma_{\cal A}$, we obtain a set of $2N_{\cal A}$ pairs of eigenvalues as
\begin{align}
&Q_{\cal A}^\dagger (i\Gamma) Q_{\cal A}= {\rm diag}\big(\,\{\pm \nu_m \}\,\big)_{m=1}^{2N_{\cal A}}, 
\quad 0\le \nu_m \le 1, \notag\\
&  (\tilde c_1, \cdots, \tilde c_{2m-1}, \tilde c_{2m},\cdots, 
 \tilde c_{2N_{\cal A}})
 \notag\\
 = &Q_{\cal A} (c_1,\cdots,  c_{2m-1},  c_{2m}, \cdots c_{2N_{\cal A}}).
\end{align}
They are put back to the fermionic operator in region ${\cal A}$ as 
$\tilde b_m=(\tilde c_{2m-1} + i \tilde c_{2m})/2$, and its number is described by the eigenvalues as 
$\langle \tilde b_m^\dagger \tilde b_m\rangle=(1+\nu_m)/2$. 
Following Ref.[\onlinecite{Vidal2003}], 
the Schmidt value is obtained for each set $\ell \equiv \{n_m^{\cal A}\}_{m=1}^{2N_{\cal A}}$ as 
\begin{align}
\lambda_\ell = \prod_{m=1}^{2N_{\cal A}} \frac{1+(-)^{n_m^{\cal A}}\nu_m }{2}. 
\label{eq:lambda-parton}
\end{align}
The important notice is that each Schmidt state $|\psi_\ell^{\cal A}\rangle$ labeled by 
$\ell=1,\cdots,2N_{\cal A}$, is the eigenstate of $\hat D_{\cal A}$ because, 
\begin{align}
\hat D_{\cal A}&=
\prod_{(\ell,s)\in {\cal A}} c_{\ell s \uparrow;1} c_{\ell s \uparrow;2} c_{\ell s \downarrow;1}c_{\ell s \downarrow;2}
\notag \\
& \quad \rightarrow\;\; D_{\cal A}=  {\rm det} Q_{\cal A} \prod_{m=1}^{2N_{\cal A}} (-1)^{n_m^{\cal A}}. 
\label{eq:DA-parton}
\end{align}
This allows us to evaluate $D_{\cal A}$ for each Schmidt state. 
\par
\begin{figure*}[tbp]
    \centering
    \includegraphics[width=17.5cm]{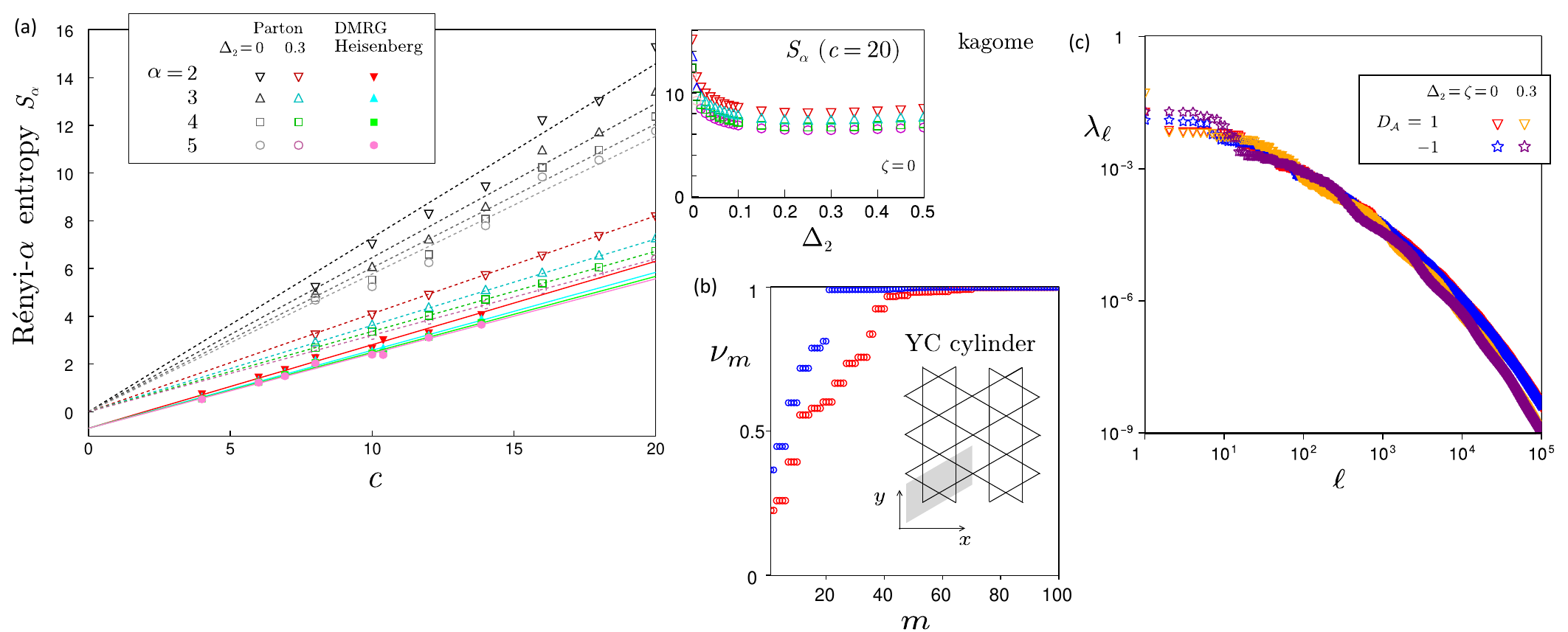}
    \caption{Results of the parton Hamiltonian on a kagome lattice with $\chi_{ij}=\pm \chi_\gamma$, 
   $\chi_1=1, \chi_2=-0.0186$, on a YC cylinder (PBC in the $y$-direction, OBC in the $x$-direction), 
   where the circumference is $c=2N_y$. We set $N_x=32$ and $N_y=4-10$. 
   (a) R\'enyi-$\alpha$ entropy $S_\alpha$ with $\alpha=2,3,4,5$, for which the previous DMRG result on the 
    Heisenberg model \cite{Schollwock2012} is available for comparison. 
     Broken lines (top five data) are the U(1) case ($\Delta_2=0$, $\zeta=0$), and the next five are the ${\mathbb Z}_2$ 
   case ($\Delta_2=0.3$, $\zeta=0$), which are different by a factor 2. 
   The DMRG data \cite{Schollwock2012} develops almost parallelly with the ${\mathbb Z}_2$ parton case 
   while exhibiting a $-\ln 2$ intercept. 
    Inset: $\Delta_2$ dependence of $S_\alpha$ at $c=20$ at $\zeta=0$. 
    (b) Eigenvalues of correlation matrix $\nu_m$ for the bipartition of the cylinder with $N_{\cal A}=960$. 
    Red circles are the U(1) case ($\Delta_2=\zeta=0$) and blue circles are the ${\mathbb Z}_2$ case ($\Delta_2=\zeta=0.3$).
    (c) Schmidt values obtained for $N_{\cal A}= 12$ taken as two unit cells of aspect ratio 1 at the center of the cylinder for  U(1) case ($\Delta_2=\zeta=0$) and the ${\mathbb Z}_2$ case ($\Delta_2=\zeta=0.3$)}, 
     where we classify them by $D_{\cal A}=\pm 1$. 
    \label{fscylinder}
\end{figure*}
\begin{figure*}[tbp]
    \centering
    \includegraphics[width=16cm]{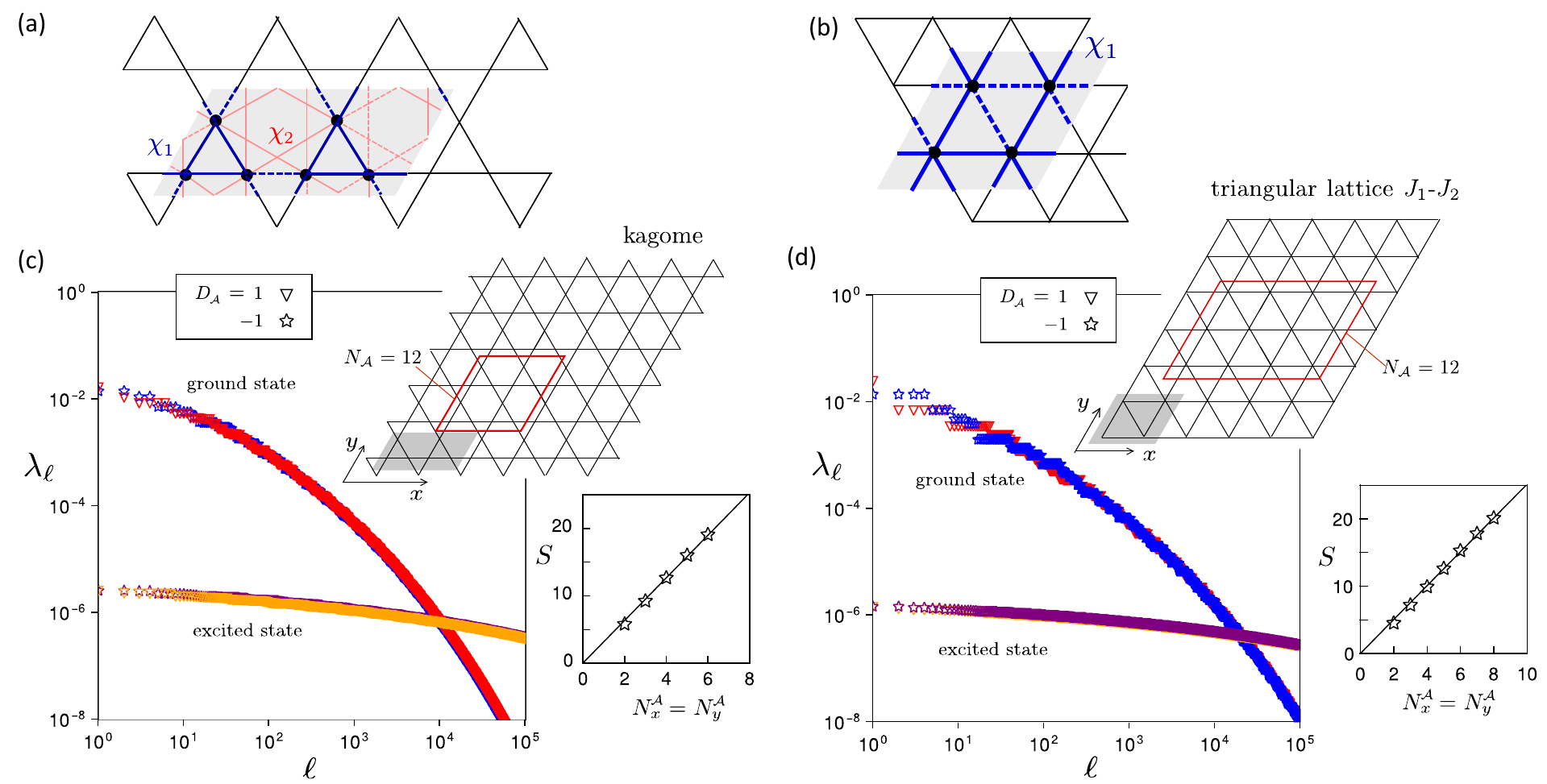}
    \caption{Results of the parton Hamiltonian on kagome and triangular lattices. 
    (a,b) Unit cell ($n_c=6$ and $4$) for the parton Hamiltonian; 
    kagome lattice has nearest neighbor $\chi_{ij}=s_{ij} \chi_1$, and next nearest neighbor $s_{ij}\chi_2$, 
    with $(\chi_1,\chi_2)=(1,-0.0186)$, where solid and broken lines indicate $s_{ij}=1$ and $-1$, respectively. 
    The triangular lattice has nearest neighbor $\chi_{ij}=s_{ij}\chi_1=\pm 1$ with $n_s=4$. 
    We set $\Delta_{ij}=\zeta=0$.
    (c,d) Schmidt spectrum obtained for $N=6N_xN_y=384$, $N_x=N_y=8$ (kagome) 
     and $N=4N_xN_y=400$ with $N_x=N_y=10$ (triangle). 
     The lattice has PBC in both $x$ and $y$ directions shown in the figure, and subsystem ${\cal A}$ is taken 
    as $N_{\cal A}=12$ ones indicated in the inset. 
    The size scaling of the entanglement entropy obtained by this analysis is given in the inset for $N_x^{\cal A}=N_y^{\cal A}$ 
    where the units of $N_x^{\cal A}$ and $N_y^{\cal A}$ are taken as those shown as shaded regions of the lattice. 
    }
    \label{fsparton}
\end{figure*}

When the U(1) symmetry is kept, the correlation matrix is block diagonalized into up and down spin sectors, 
and thus the Schmidt values become the product of the two spin sectors as 
\begin{align}
\lambda_\ell = \prod_{m=1}^{N_{\cal A}} \frac{1+(-)^{n_{m\uparrow}}\nu_m }{2}
\prod_{m'=1}^{N_{\cal A}} \frac{1+(-)^{n_{m'\downarrow}}\nu_{m'} }{2}. 
\end{align}
This fact indicates that the Schmidt states are completely separable between the up and down spin components, 
namely, the two components have zero entanglement. 
In that case, the entanglement entropy is determined (as in the case of standard free fermions) 
by counting the probability of each one-body fermionic eigenstate having the distributions on both sides. 
When the U(1) symmetry is broken and the gap opens, the ${\mathbb Z}_2$ spin liquid state is obtained. 
This state is analogous to the superconductivity of the Kitaev chain, in which case the fermions form a Cooper pair. 
Their entanglement entropy is obtained by counting the probability distribution of each Cooper-pair eigenstate, 
which number is roughly half the original fermionic number (see Fig.~\ref{fscylinder}(b)). 
This explains why the introduction of $\Delta_2 \ne 0$ reduces the entanglement entropy by half from the U(1) case 
in Fig.~\ref{f4}(a) in the main text (see also Fig.\ref{fscylinder}(a)). 

%
%
\subsection{Pfaffian and U(1) and ${\mathbb Z}_2$ Spin liquids}
In this subsection, we explain in detail how the definition of $D$ in Eq.(\ref{eq:z2eq}) in the main text is related 
to the characterization of the spin liquid. 
In the same way as Eq.(\ref{eq:z2eq}) in the main text, the global operator $\hat D$ is given in the form, 
\begin{align}
\hat D&=\prod_{(\ell,s)} c_{\ell s \uparrow;1} c_{l s \uparrow;2} c_{\ell s \downarrow;1}c_{\ell s \downarrow;2}
\notag \\
& \quad \rightarrow\;\; D= {\rm det} Q \prod_{m=1}^{2N} (-1)^{n_m} 
={\rm sgn}({\rm Pf}(A)) (-)^{\sum_m n_m} , 
\label{eq:DA-2}
\end{align}
where $n_m=0,1$ is the number of fermions occupying the $m$-th eigenstate of Eq.(\ref{eq:BdG}), 
$A$ is an antisymmetrc matrix defined as ${\cal H}_p = \frac{i}{4}\sum_{\ell s \sigma; \alpha,\;\ell' s' \sigma'; \alpha'} A_{\ell s \sigma; \alpha,\;\ell' s' \sigma'; \alpha'} c_{\ell s \sigma; \alpha} c_{\ell' s' \sigma'; \alpha'}$
and ${\rm det}Q= (-)^{2N} \sqrt{{\rm det} A}/{\rm Pf}(A)= {\rm sgn}({\rm Pf}(A))$. 
\par
Let us make some remarks on the relationship between $D$ and the Pfaffian. 
When $\Delta_{ij}=\zeta_{ij}=0$ with unbroken U(1) symmetry, the up and down-spin fermions are decoupled, 
so that $Q^\uparrow=Q^\downarrow$. 
The Bogoliubov fermion number for spin $\sigma$ is defined as $n_{m\sigma}$, with 
$n_{m}=n_{m\uparrow}+n_{m\downarrow}$. Then, $D$ reduces to 
\begin{align}
\notag\\
& D={\rm det} Q^\uparrow {\rm det} Q^\downarrow \prod_{m=1}^{N} (-1)^{n_{m \uparrow }+ n_{m\downarrow}} 
  =\prod_{m=1}^{N} (-1)^{n_{m \uparrow }+ n_{m\downarrow}}, 
\label{eq:DA-U1}
\end{align}
which means that $D$ no longer depends on the Pfaffian, 
as we always find ${\rm det}Q={\rm det} Q^\uparrow {\rm det} Q^\downarrow=({\rm det} Q^\downarrow)^2=1$. 
\par
The characterization of the U(1) spin liquid is given by Wen\cite{Wen2002} using a Wilson loop operator  
along the closed loop $C$ ;  
Using a matrix $U_{ij}$ defined on an $\langle ij\rangle$ neighboring bond, 
the matrix representation of the Wilson loop is given as
\begin{equation}
U_{ij}= \left(\begin{array}{rr}
\chi_{ij} & \Delta_{ij} \\
\Delta_{ij}^* & -\chi_{ij}
\end{array}
\right), \quad\quad
{\cal P}_C= \prod_{\langle i,j\rangle\in C} U_{ij}.
\end{equation}
When $\Delta_{ij}=0$, $U_{ij}$ is diagonal and accordingly, Wilson loop commutes, $[P_C, P_{C'}]=0$. 
In this case, $\chi_{ij}\ne 0$ yields a U(1) spin liquid. 
\par
When $\Delta_{ij}\ne 0$, the noncommutative loops, $[P_C, P_{C'}]\ne 0$, yield a ${\mathbb Z}_2$ spin liquid. 
This means that from Eq.(\ref{eq:DA-2}), we can find ${\rm det}Q \ne 1$ only when the U(1) symmetry is broken, 
in which case ${\rm sgn(Pf}(A))$ enters $D$. 


\subsection{Numerical results}
\label{app:partonnumerical}
We now demonstrate the validity of applying parton theory to the $S=1/2$ kagome and triangular lattice $J_1$-$J_2$ Heisenberg models. 
For the kagome lattice, we adopted $\chi_{ij}=\pm \chi_1$ for the nearest neighbor bonds 
and $\chi_{ij}=\pm \chi_2$ for the next-nearest neighbor bonds 
in a configuration shown in Fig.~\ref{fsparton}(a) with six-site unit cell. 
The plus and minus bonds are shown in solid and broken lines, respectively, 
corresponding to the ${\mathbb Z}_2[0,\pi]\beta$ ansatz. 
The parameter values are taken from the variational Monte Carlo study (VMC) in Ref.[\onlinecite{Iqbal2011}] as 
$(\chi_1,\chi_2)=(1,-0.0186)$. 
For a U(1) spin liquid, $\Delta_2=\zeta=0$, and for the ${\mathbb Z}_2$ spin liquid, 
we choose $\Delta_2, \zeta \ne 0$ by hand. 
In the VMC studies \cite{Iqbal2011,Iqbal2013}, the optimization of parameters 
reach $\Delta_2=\zeta=0$, while their convergence is not as good as the energy itself; 
the free energy landscape about $\Delta_2$, 
that connects U(1) and ${\mathbb Z}_2[0,\pi]\beta$, is reported to be very flat. 
This indicates the actual difficulty of determining which is stable, solely by numerical simulations, 
as has been discussed for decades\cite{Lauchli2019}. 
Here, we do not aim to discuss this issue, but instead, 
make direct comparison of the entanglement property between 
the parton spin liquid state and the ${\mathbb Z}_2$ spin liquid reported previously in the DMRG study\cite{Schollwock2012}. 
\par
We first show the results of the kagome lattice calculated on a series of YC cylinders
shown in the inset of Fig.~\ref{fscylinder}; 
the periodic boundary (PBC) taken in the $y$-direction, and its circumference is $c=2N_y$, 
where we set $N_x=32$ and $N_y=4-10$, with $N=6N_xN_y$ up to 1920 sites. 
We show in Fig.~\ref{fscylinder}(a), the R\'enyi-$\alpha$ entropy defined as 
\begin{align}
S_\alpha^f = \frac{1}{1-\alpha} \ln {\rm Tr}_A[(\rho_A)^\alpha], 
\end{align}
to compare with those obtained by the cylinder DMRG in Ref.[\onlinecite{Schollwock2012}]. 
We plot $\alpha=2,3,4,5$ for the U(1) parton state, 
where $\alpha=2$ data is the same as the one plotted in Fig.~\ref{f4}(a) in the main text. 
As we show in the inset for the $c=20$ case, 
breaking the U(1) symmetry down to ${\mathbb Z}_2$ by the introduction of $\Delta_2\ne 0$ 
reduces $S_\alpha$ by half. 
Then, we find that the slope of $S_\alpha/2$ of our Parton Hamiltonian (fermions) 
for all $\alpha$ in the main panel is comparable to the one obtained 
by DMRG in the $S=1/2$ kagome Heisenberg antiferromagnet (see Fig.~6 of Ref.[\onlinecite{Schollwock2012}]). 
The difference is the emergent $-\ln 2$; it means that the $D_{\cal A}=1$ sector of the parton state has 
quantitatively the same entanglement scaling law as the spin liquid state in DMRG, 
and that the Gutzwiller projection is the origin of the $-\ln 2$ term. 
\par
To examine how the two $D_{\cal A}$ sectors emerge, we plot in Fig.~\ref{fscylinder}(b) 
the distribution of the eigenvalues of the correlation matrix, $\nu_m$, which generates $\lambda_\ell$ 
according to Eq.(\ref{eq:lambda-parton}). 
Here, naively, $\nu_m$ is considered as an occupation number of fermionic particles.  
By comparing the case of $\Delta_2=\zeta=0$ (U(1)) and $0.3$ (${\mathbb Z}_2$), 
the occupation is roughly reduced by half, which is intuitively understood as the aforementioned 
analogy with the Kitaev superconductor, that the pairing of fermions reduces the occupation number by half. 
In fact, as we showed in the main text, introducing $\Delta_2=\zeta\ne 0$ will almost perfectly 
support the half-reduction of the entanglement entropy due to the breaking of U(1) symmetry. 
We then further show in Fig.~\ref{fscylinder}(c) the Schmidt values, $\lambda_\ell$, 
for the U(1) and ${\mathbb Z}_2$ states. 
The Schmidt spectrum splits into two nearly degenerate $D_{\cal A}=\pm 1$ sectors, 
and by definition, the Gutzwiller projection removes the $-1$ sector to put the state back to the 
spin representation, where the $-\ln 2$ appears. 
\par
Figure~\ref{fsparton}(c) shows the results of the parton theory for the same kagome lattice antiferromagnet 
on a torus (PBC in two different directions shown in the inset), 
where the unit cell is the shaded $n_c=6$ sites. 
We calculated the Schmidt values for subsystem ${\cal A}$ consisting of $N_{\cal A}=12$ sites shown in the panel; 
they are separated into almost equivalent $D_{\cal A}=\pm 1$ sectors, 
where the finite-size (shape) effects are less pronounced than in the cylinders. 
The excited state has the same feature with more precise degeneracies. 
The size scaling of the entanglement entropy as a function of  $N_{{\cal A}x}=N_{{\cal A}y}$ 
is shown in the inset, which linearly extrapolates to zero. 
By removing the $D_{\cal A}=-1$ contribution from it, we obtain the $-\ln 2$ term. 
\par
We also performed the same calculation for the triangular lattice $J_1$-$J_2$ antiferromagnet \cite{Iqbal2016} shown in Fig.~\ref{fsparton}(b), 
where we set the nearest neighbor $\chi_{ij}=\pm 1$ as solid and broken lines with 
$n_s=4$ site unit cell, which has a $\pi$ flux on the upward triangle. 
By setting $N_x=N_y=10$ with $N=4N_xN_y=400$, and the subsystem ${\cal A}$ as $N_{\cal A}=12$ shown in 
Fig.~\ref{fsparton}(d), we find results similar to the kagome lattice case. 
The size scaling for $N_{x}^{\cal A}=N_{y}^{\cal A}$ shows a linear slope toward zero, which we can expect to acquire $-\ln2$ after 
Gutzwiller projection onto the $D_{\cal A}=1$ sector.  
Unfortunately, unlike the kagome lattice, there are no rigorous numerical simulations 
that exhibit the entanglement entropy (or R\'enyi entropy) scaling, which makes the direct comparison difficult. 
However, the features observed here for the parton Hamiltonian are basically equivalent to the kagome lattice. 
This fact strongly supports the general picture of the reduction of Schmidt states as 
the cause of the topological term. 

\bibliography{kitaev_ref}

\end{document}